## An Experimental Investigation of Cavitation Bulk Nanobubbles Characteristics: Effects of pH and Surface-active Agents

SCHOLARONE™
Manuscripts

# An Experimental Investigation of Cavitation Bulk Nanobubbles Characteristics: Effects of pH and Surface-active Agents

Ritesh Prakash[1], Jinseok Lee[2], Youngkwang Moon[2], Diva Pradhan[2], Seung-Hyun Kim[3], Ho-Yong Lee[4], and Jinkee Lee[1, 2]*

[1] Microfluidic Convergence Laboratory, Institute of Quantum Biophysics, Sungkyunkwan University, Suwon 16419, Republic of Korea

[2] Microfluidic Convergence Laboratory, School of Mechanical Engineering, Sungkyunkwan University, Suwon 16419, Republic of Korea

[3] School of Engineering, Brown University, Providence, Rhode Island 02912, USA

[4] Department of Material Science and Engineering, Sunmoon University, Asan 31460, Republic of Korea

*Corresponding author: Prof. Jinkee Lee, lee.jinkee@skku.edu

## Abstract

Understanding the behavior of nanobubbles (NBs) in various aqueous solutions is a challenging task. The present work investigates the effects of various surfactants (*i.e.*, anionic, cationic, and nonionic) and pH medium on bulk NBs formation, size, concentration, bubble size distribution (BSD), zeta potential, and stability. The effect of surfactant was investigated at various concentrations above and below critical micelle concentrations. NBs were created in DI water using a piezoelectric transducer. The stability of NBs was assessed by tracking the change in size and concentration over time. NBs size is small in the neutral medium compared to the other surfactant or pH mediums. The size, concentration, BSD, and stability of NBs are strongly influenced by the zeta potential rather than the solution medium. BSD curve shifts to lower bubble sizes when the magnitude of zeta potential is high in any solution. NBs were observed to exist for a long time, either in pure water, surfactant, or pH solutions. The longevity of NBs is shortened in environments with pH $\leq 3$. Surfactant adsorption on the NBs surface increases with surfactant concentration up to a certain limit, beyond which it declines considerably. The Derjaguin-Landau-Verwey-Overbeek (DLVO) theory was used to interpret the NBs stability, which resulted in a total potential energy barrier that is positive and greater than $43.90 k_B T$ for pH ranging from 6.0 to 11.0, whereas, for pH below 6, the potential energy barrier essentially vanishes. The present research will extend the in-depth investigation of NBs for industrial applications involving NBs.

## 1. Introduction

Bubbles with a size of less than 1 $\mu$m are known as nanobubbles (NBs). NBs have piqued the interest of researchers because of their physicochemical properties and wide-ranging applications.[1] NBs have widespread applications in drug delivery,[2] flotations,[3,4] cleaning,[5,6] plant growth,[7] bactericidal activity,[8] nanomedicine,[9] protein adsorption,[10] friction modifier,[11] food processing,[12] wastewater treatment,[13] as well as an ultrasound contrast agent.[14,15] In most of these applications, process efficiency depends on the properties of NBs suspensions as determined by their size, bubble size distribution (BSD), stability, and surface charge. NBs occur in two different forms, bulk NBs and surface NBs. The first one is suspended in the medium, while the latter is trapped close to the surfaces.[16,17] In this work, the discussion will be restricted to only bulk NBs. This work mainly focuses on understanding the NBs characteristics (*i.e.*, size, BSD, and surfactant adsorption) and their stability at different pHs (*i.e.*, acidic and basic) and surfactant types (*i.e.*, cationic, anionic, and nonionic) with its concentration. Literature lacks studies on NBs characteristics in the presence of surfactants. Further investigations are required to interpret the underlying science in the field.

The most peculiar characteristics of NBs are their small size, low buoyancy, and longevity. NBs exhibit enigmatic long-term stability, despite the fact that the gas diffusion theory predicts that the bubbles with a diameter of between 100 nm and 200 nm will have a very short lifetime of 20 to 80 $\mu$s.[18,19] Due to its very small size, according to the Young-Laplace equation, the internal pressure is roughly 15 atm.[20] The rate of dissolution of NBs increases as the bubble shrinks, which is attributable to increasing Laplace internal pressure, $\Delta P = 2\gamma / r$, where $\gamma$ is the surface tension and $r$ is the bubble radius. Consequently, nanoscopic bubbles cannot endure for a long time. In contrast to expectations, NBs can survive for days, weeks, and even months.[21,22] There may be several factors that can confer stability. Henry *et al.* [23] found that the buoyancy force on NBs is extremely low, with a rising velocity of 20 − 30 nm/s for a bubble with a diameter of 100 nm. A low buoyancy force could slow down the bubble dissipation rate. Other factors are the electrostatic double-layer repulsion and electrostatic pressure surrounding the bubble because of the charged bubble surface.[24] Electrostatic pressure balances the internal Laplace pressure, thus reducing and preventing the probability of NBs dissolution. Based on the aforementioned literature, it can be said that the understanding of the physical behavior of NBs is complicated, and the mechanism of the longevity of NBs in different liquid mediums is still debatable and a matter of research.

With the advancement in the visualization technique, several evidences of the existence of NBs are addressed. Numerous techniques are available to generate the bubbles at the nanoscale. These techniques are electrolysis,[25,26] sonication causing acoustic cavitation,[27] pore pressure through the porous membrane,[8,28] hydrodynamic cavitation,[29] microfluidic devices,[30] *etc*. One of these techniques is the cavitation of NBs by ultrasonication, in which local pressure in the expansion phase of the cycle falls below the vapor pressure of the liquid, causing the tiny bubble to grow. These bubbles are generated from existing gas nuclei within the fluid. The techniques for confirmation and identification of NBs are transmission electron microscope (TEM),[30,31] cryogenic electron microscopy (Cryo-EM),[32,33] dynamic light scattering (DLS),[34] nanoparticle tracking analysis (NTA),[22,27] and resonant mass measurement.[35] The NTA is the easiest and most reliable for bulk suspension among these techniques. Ma *et al*.[27] used both DLS and NTA to quantify the bubble diameter of the same sample. According to their findings, DLS overestimates the bubble diameter by 50 nm compared to the NTA. Another study also showed that the NTA of NBs is more reliable compared to the DLS, as the DLS is biased towards large bubble sizes.[21]

The degree of colloidal stability of bubbles at the nanoscale is concomitant with the magnitude of the absolute value of zeta potential. Zeta potential is significantly affected by the pH change and the presence of surfactants. Several investigators elucidated the increase in NBs size with a decrease in pH.[36,37] The NBs number density was also noticed to be reduced with a reduction in pH. Surfactants play an influential role in the nucleation, physicochemical properties, and stability of NBs. Lee and Kim[38] investigated the NBs concentration and size in the SDS solution. They elucidated a high bubble concentration and small NBs in the surfactant solution. In contrast, Nirmalkar *et al*.[36] quantified that the presence of surfactants only enhances the stability rather than affects the size and concentration in the presence of anionic surfactants. The cationic surfactant tends to neutralize the surface potential of NBs, resulting in charge reversal at the interface. In the presence of nonionic surfactants, the colloidal suspension is endowed with steric stabilization rather than affecting the size and concentration of NBs.[36] Xiao *et al*.[39] used molecular dynamics simulation to demonstrate that NBs are unstable in surfactants. Two surfactant-induced molecular mechanisms are involved in the instability of NBs: depinning a contact line and reducing vapor-liquid surface tension. The first mechanism is explicated by the considerable adsorption of surfactants on the surface, which causes depinning of the NBs contact line and, therefore, the NBs instability. The latter emphasizes surfactant adsorption at the gas-liquid interface of NBs, which results in a decrease in surface tension and an irreversible liquid-to-vapor phase shift. The influence of surfactants on NBs properties is unclear in the literature, and no general hypothesis is recognized to explain NBs stability.

Several investigators have suggested the existence of an electric double layer (EDL) surrounding the NB in pure water.[36,40,41] The inter-colloidal interaction between the NBs can be well explained by the DLVO theory. This theory is valid by assuming that the nanosized entity is at infinite dilution. The total interaction energy at the nanoscale will be the sum of van der Waals and electrostatic repulsive force.[42] Electrostatic repulsive forces must be dominant in the suspension to maintain evenly and stably suspended bubbles. Some important interaction forces acting between two identical nano entities in the aqueous media are the electric double layer, steric interaction, van der Waals, and hydrophobic interaction. The first two are repulsive, whereas the remaining are attractive in nature.

The potential of NBs in industrial applications has piqued the great attention of researchers around the world. Despite significant advances in NBs research, it is observed that there are limited comprehensive investigations on the role of surfactants on NBs surface charge, size, BSD, concentration, and longevity available in the literature. Therefore, the present research work aims to characterize the NBs in the presence of three different types of surfactants, *i.e.*, cationic, anionic, and nonionic, by focusing on their stability mechanisms and potential reasons. Moreover, the influence of pH is also studied. In addition, surfactant adsorption and its dissociation characteristics on the NBs surface in the presence of an ionic surfactant are also enunciated. Also, published literature does not elucidate the distribution of ions and electric double layer around the NBs in the presence of ionic surfactants. Furthermore, an attempt has also been made to explicate the plausible prospect of ion distribution and its alignment surrounding NBs in cationic and anionic surfactants.

## 2. Theoretical Background

## 2.1. DLVO theory to interpret NBs stability

A colloidal system is always subjected to Brownian motion, with frequent collisions between colloidal entities. The interactions between the entities in such collisions determine the stability behavior. The suspension may become unstable if the attraction between entities is too strong, while the system will remain in a dispersed condition if repulsive forces predominate. Interpretation of the stability of NBs can be enunciated by DLVO theory. As per extended DLVO theory, van der Waals interaction potential, electrostatic repulsive interaction potential, and hydrophobic interaction potential are responsible for the stability of the colloidal suspension. Dipolar interaction between the assemblies of molecules in each nano entity is responsible for van der Waals interaction, whereas electrostatic repulsive interaction is caused by the same-charged electrical double layer surrounding NBs. Israelachvili and Pashley [43] first introduced the concept of hydrophobic forces. Hydrophobic interaction potential can occur naturally or be induced by the adsorbed hydrophobic species.[44] Hydrophobic interactions are found to have a far greater range than the van der Waals interaction potential.[45] Total interaction between two NBs can be represented as[46-49]

$$E_{\mathrm{T}} = E_{\mathrm{A}} + E_{\mathrm{H}} + E_{\mathrm{R}} \tag{1}$$

where $E_{\mathrm{T}}$, $E_{\mathrm{A}}$, $E_{\mathrm{H}}$, and $E_{\mathrm{R}}$ refer to total interaction potential, van der Waals interaction potential (attractive), hydrophobic interaction potential, and electrostatic repulsive interaction energy (repulsive). Total potential energy can be normalized by the microscopic thermal energy of the molecule $(k_{\mathrm{B}}T$, where $k_{\mathrm{B}} = 1.380 \times 10^{-23}$ J/K and $T = 298.15$ K$)$. Equation (1) signifies the variation in total interaction potential as two NBs approach each other, indicating whether they experience attraction, repulsion, or hydrophobic interaction. In Eq. (1), $E_{\mathrm{A}} + E_{\mathrm{H}}$ can be represented as

$$E_{\mathrm{A}} + E_{\mathrm{H}} = \frac{-\left(A_{\mathrm{H}} + K\right)}{6}\left[\frac{2r_1 r_2}{H^2 - \left(r_1 + r_2\right)^2} + \frac{2r_1 r_2}{H^2 - \left(r_1 - r_2\right)^2} + \ln\left(\frac{H^2 - \left(r_1 + r_2\right)^2}{H^2 - \left(r_1 - r_2\right)^2}\right)\right] \tag{2}$$

where $A_{\mathrm{H}}$ denotes the Hamker constant, $K$ is the hydrophobic constant, and $r_1$ and $r_2$ is the NBs radius. Another researcher attempted to deduce the $K$ from the DLVO theory.[50] The importance of the size effect on colloidal system stability is demonstrated by Eq. (2). The value of $K$ for SDS surfactant in the presence of different concentrations of NaCl salt was $10^{-19}$ J, whereas it was $10^{-17}$ J in the absence of salt. In this work, the value of $K$ is taken as $10^{-19}$ J.[49] The term $H$ refers to the center-to-center distance between two NBs, which is defined as $H = r_1 + r_2 + d$. Here, $d$ denotes the separation distance between two NBs. Based on the Lifshitz theory, the Hamker constant for the air-water system may be obtained as

$$A_{\mathrm{H}} = 0.75 k_{\mathrm{B}} T \left( \frac{\varepsilon - \varepsilon_0}{\varepsilon - \varepsilon_0} \right) + \frac{3 h \nu_{\mathrm{e}}}{16\sqrt{2}} \frac{\left(n_1^2 - n_2^2\right)^2}{\left(n_1^2 + n_2^2\right)^{1.5}} \tag{3}$$

where $h$ is the Planck constant, $\varepsilon$ is the permittivity of the dispersed phase, $\varepsilon_0$ is the permittivity of the vacuum, $\nu_{\mathrm{e}}$ is the water absorption frequency, $T$ is the temperature, $k_{\mathrm{B}}$ is the Boltzmann constant, $n_1$ is the refractive index of air, and $n_2$ is the refractive index of the medium. The values of these are: $h = 6.626 \times 10^{-34}$ J, $\varepsilon = 8.85 \times 10^{-12}$ F/m, $\varepsilon_0 = 78.4$, $\nu_{\mathrm{e}} = 3.0 \times 10^{-15}$ 1/s, $n_1 = 1.0$, and $n_2 = 1.33$. Hamker constant is an inherent property of materials, and its value indicates the strength of van der Waals interactions.[51] Electrostatic repulsive interaction energy can be calculated as

$$E_{\mathrm{R}} = \frac{\pi \varepsilon \varepsilon_0 r_1 r_2 \left(\psi_1^2 + \psi_2^2\right)}{r_1 + r_2} \left[ \frac{2\psi_1 \psi_2}{\psi_1^2 + \psi_{21}^2} \ln\left( \frac{1 + \exp(-kh)}{1 - \exp(-kh)} \right) + \ln\left(1 - \exp(-2kh)\right) \right] \tag{4}$$

where $\psi_1$ and $\psi_2$ refer to the surface potentials of NBs of radius $r_1$, and $r_2$, respectively. NBs contain surface charge and form an electric double layer attributed to the existence of the co- and counterions. The thickness of the double layer surrounding the charged NBs is Debye length $(k^{-1})$, which can be expressed as

$$k^{-1} = \sqrt{\frac{\varepsilon \varepsilon_0 k_{\mathrm{B}} T}{2 N_{\mathrm{A}} I e^2}} \tag{5}$$

where $N_{\mathrm{A}}$ refers to the Avogadro's number, $I$ is the ionic strength of the solution, and $e$ is the electronic charge of the proton. The values are: $N_{\mathrm{A}} = 6.023 \times 10^{23}$ molecules/mol, $e = 1.60 \times 10^{-19}$ C. The value of $k^{-1}$ in pure water is reported to be 961 nm.[41]

## 2.2. Adsorption and surface ionization of surfactants at NBs surface

Understanding the gas-liquid interfacial characteristics of NBs in the presence of an ionic surfactant is essential for interpreting the behavior of colloidal suspension. The surface characteristics of NBs depend on the surfactant concentration and length of their alkyl chains.[52] The surface charge of the NBs governs the adsorption of ionic surfactants. The adsorption of ionic surfactant on the NBs surface may form a double layer, and the magnitude of charge at the double layer is characterized by the concentration and charge of the surfactant monomers. Ionic surfactants are amphiphiles with polar and non-polar groups. Due to the differing orientations of polar and non-polar groups, an electric double layer forms at the gas-liquid interface. The formation of a double layer around the bubble precludes the further adsorption of surfactant monomers on the bubble surface. However, it facilitates the attraction of counterions. The presence of ionic surfactants at the gas-liquid interface crucially impacts the NBs behavior. In bulk solutions, surfactants are completely ionized, but the amount of ions formed at the NBs interface is constrained by the degree of dissociation. The adsorption of ionic surfactants on the NBs surface can be analyzed using the Gibbs adsorption expression[53,54]

$$\Gamma_s = \frac{-1}{nRT} \frac{d\gamma}{d \ln c_s} \qquad (6)$$

where $\Gamma_s$ denotes the Gibbs surface excess in mol/m$^2$, $R$ refers to the gas constant (8.314 J/mol.K), and $c_s$ is the surfactant concentration in mM. The value of $n$ can be taken as 1 for nonionic surfactants and $n = 2$ or 3 for mono or divalent counterions.[54] In this study, $n = 2$ was employed for ionic surfactants.[55] The slope $d\gamma / d \ln c_s$ in Eq. (6) can be obtained from the plot of $\gamma$ against $\ln c_s$. The magnitude of the slope was found to increase with an increase in the surfactant concentration. The larger the −ive value of the slope, the higher the surfactant adsorption. Gibbs surface excess values can be used to analyze the number of adsorbed molecules per unit area. The number of adsorbed molecules per unit area can be determined as $\Gamma_s N_A$. Assuming complete dissociation of surfactant at the NBs surface, the surface charge density $(\sigma_0)$ due to the amount of ionized surfactants can be estimated by[56]

$$\sigma_0 = e N_A \Gamma_s \qquad (7)$$

Surface charge density as a function of surface potential $(\psi_0)$ can also be determined using Eq. (8)[57]

$$\sigma_0 = z\sqrt{2\varepsilon\varepsilon_0 k_B T \rho_{bulk}} \left[ 2\sinh\left(\frac{e\psi_0}{2k_B T}\right) + \frac{4}{r}\sqrt{\frac{\varepsilon\varepsilon_0 k_B T}{2e^2 z^2 \rho_{bulk}}}\tanh\left(\frac{e\psi_0}{4k_B T}\right) \right] \tag{8}$$

where $z$ denotes the ionic valence, $\rho_{bulk}$ is the concentration of the ions in bulk. The surface potential value can be quantified using Eqs. (7) and (8). Some counterions may penetrate the stern layer; hence the net charge density $(\sigma)$ in the stern layer can be given as[49]

$$\sigma = \sigma_0 - \sigma_c = eN_A\left(\Gamma_s - \Gamma_c\right) \tag{9}$$

where $\sigma_c$ is the surface charge density in the stern layer and $\Gamma_c$ is the Gibbs surface excess of counterions. The value of zeta potential, $|\zeta|$ is generally less than or equal to the stern potential $|\psi_c|$. In the case of strong ionic concentration $|\zeta| \leq |\psi_c|$, whereas in low ionic strength, $|\zeta| \cong |\psi_c|$.[58] In this work, it is assumed that the $|\psi_c| = |\zeta|$ and the stern potential can be calculated by replacing $\sigma_c$ instead of $\sigma_0$ in the left-hand side of Eq. (8). The Gibbs surface excess of the counterions in the stern layer can be interpreted as

$$\Gamma_c = \Gamma_s - \frac{\sigma}{eN_A} \tag{10}$$

Surface potential can be determined as

$$\psi_0 = \psi_c^{drop} + \zeta \tag{11}$$

where $\psi_c^{drop}$ is the potential drop in the stern layer, which can be interpreted as

$$\psi_c^{drop} = \sigma\frac{\delta}{\varepsilon\varepsilon_0} \tag{12}$$

where $\delta$ is the distance between the polar heads of the −ive and +ive ions in the stern layer. Substituting Eq. (12) into Eq. (11), the final form of the surface potential can be expressed as

$$\psi_0 = \sigma \frac{\delta}{\varepsilon \varepsilon_0} + \zeta \tag{13}$$

Total surfactant molecules $(\Gamma)$ at the surface can be determined as the sum of the adsorbed –ive surfactant head groups and +ive counterions as

$$\Gamma = \Gamma_s + \Gamma_c \tag{14}$$

Surfactant dissociation can be estimated by

$$\beta = \frac{\Gamma_s}{\Gamma} \tag{15}$$

where $\beta$ refers to the surfactant dissociation. The molecular area of surfactant in $nm^2$/molecule at the interface can be enunciated as[59-61]

$$a = \frac{10^{18}}{N_A \Gamma} \tag{16}$$

where $a$ denotes the molecular area of the adsorbed surfactant. Having the knowledge of surface charge density (calculated from Eq. 7) and surface potential (determined from Eqs. 7 and 8) in the presence of an ionic surfactant, Debye length can be quantified as[62]

$$k^{-1} = \frac{\psi_0 \varepsilon \varepsilon_0}{\sigma_0} \tag{17}$$

## 3. Experimentation

### 3. 1. Experimental Setup

All the experiments were performed at room ambient temperature and pressure. The outline of the experimental setup is shown in Figure 1. The experimental setup consists of the piezoelectric transducer, the power amplifier, the current probe, the signal generator, a computer, and a beaker. The piezoelectric transducer (Se Jeong - Hitech, Republic of Korea) was used to generate the NBs. The NBs were produced in deionized (DI) water. The signal generator software (supplied with the signal generator) was used to set the frequency and voltage of the piezoelectric transducer. The computer was connected to the signal generator (TiePie, HS3-100, Netherland), which was linked to the voltage amplifier (Matsusada, DOA150-1, Japan). Finally, the voltage amplifier was coupled to the piezoelectric transducer and operated it. The voltage amplifier amplifies the continuous sinusoidal wave generated by the signal generator. The current detection probe was connected between the signal generator and the +ive end of the piezoelectric transducer to determine the real-time power input to the piezoelectric transducer. The actual electric power delivered to transducers was determined using a voltage at both the +ive and –ive ends of the transducers, and a current probe measured the current. The experiments were conducted at a resonance frequency of 28.43 kHz, and the input voltage through the oscilloscope and function generator was 6 V. The amplified root mean square (RMS) voltage was 159.60 V. The piezoelectric transducer was fixed at the top of the beaker and its small portion descended into the liquid while producing NBs. The NBs were prepared in a square beaker with a side of 0.08 m and a height of 0.10 m. The amount of DI water was kept at 0.50 L for all the experiments.

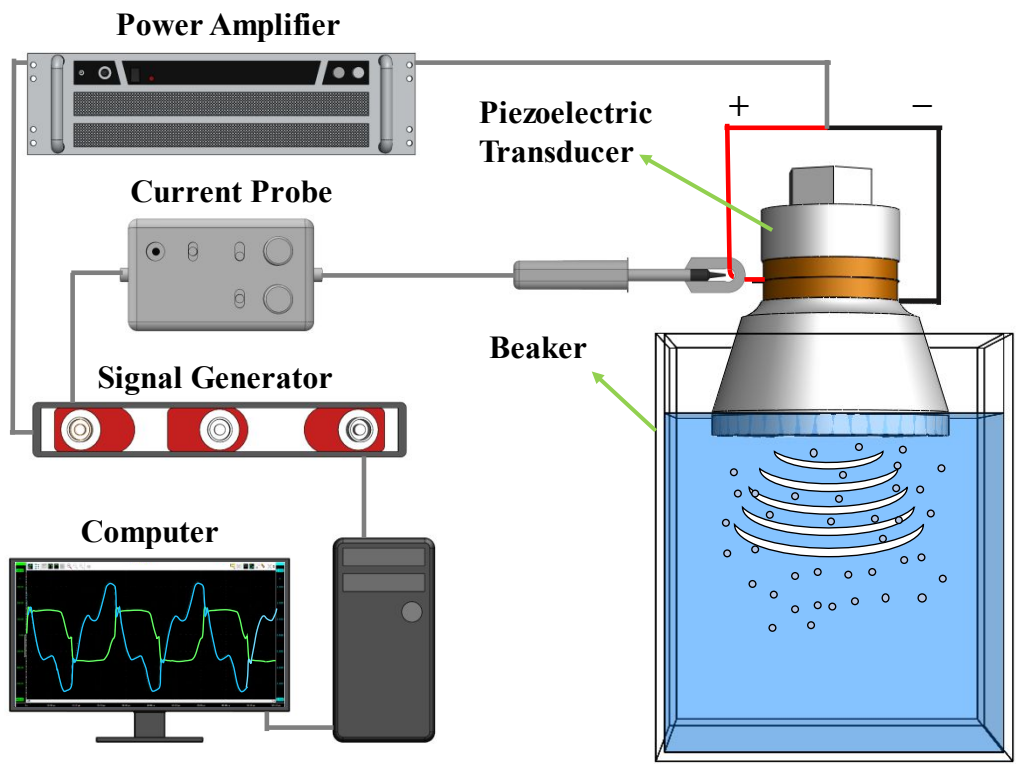

Figure 1. Schematic of the experimental arrangement.

## 3.2. Materials and Methods

The pH and electrical conductivity of DI water were 7.03 and 4.0 $\mu s$/cm. DI water was changed to the acidic medium using hydrochloric acid (HCl) (AR = 37%), while the basic medium was adjusted using sodium hydroxide (NaOH) (50% diluted, AR = 48 – 52%). Both HCl and NaOH were of analytical purity. The dropwise technique was used to change the DI water to an acidic and basic medium. The samples were thoroughly mixed while adjusting the pH using the magnetic stirrer. The pH of the samples was measured by a pH meter (Thermo Scientific, Orion Star A215). NBs were also generated using three different surfactants, such as sodium dodecyl sulfate (SDS) (AR ≥ 99%) (MW = 288.40 g/mol), hexadecyltrimethylammonium bromide (CTAB) (AR ≥ 98%, MW = 364.45 g/mol), and Tween 20 (AR = 40%, MW = 1228 g/mol). All the selected surfactants are water-soluble. All chemicals were laboratory grade and purchased from Sigma Aldrich, USA. Before generating the bulk NBs in the surfactant solution, the required quantity of surfactant was measured and dissolved in the DI water by stirring with a magnetic stirrer. The solutions were mixed for 30 min using the magnetic stirrer. In all the experiments, the suspension of NBs in different solutions was prepared for 45 minutes. After generating NBs, samples were kept in air-tight 70 mL glass vials at room temperature for further analysis. After finishing the experiments, the beaker and the surface of the piezoelectric transducer were thoroughly cleaned with DI water to avoid the contamination of the reagents or chemicals from the previous experiment. The surface tension ($\gamma$) of liquid containing surfactant was measured by a rheometer (TA Instruments, HR 30, USA) with interfacial tension measurement accessories. *i.e.*, double wall ring (DWR), geometry holder, and sample holder at 298.15 K. The surface tension against surfactant concentration results are depicted in Figure 2.

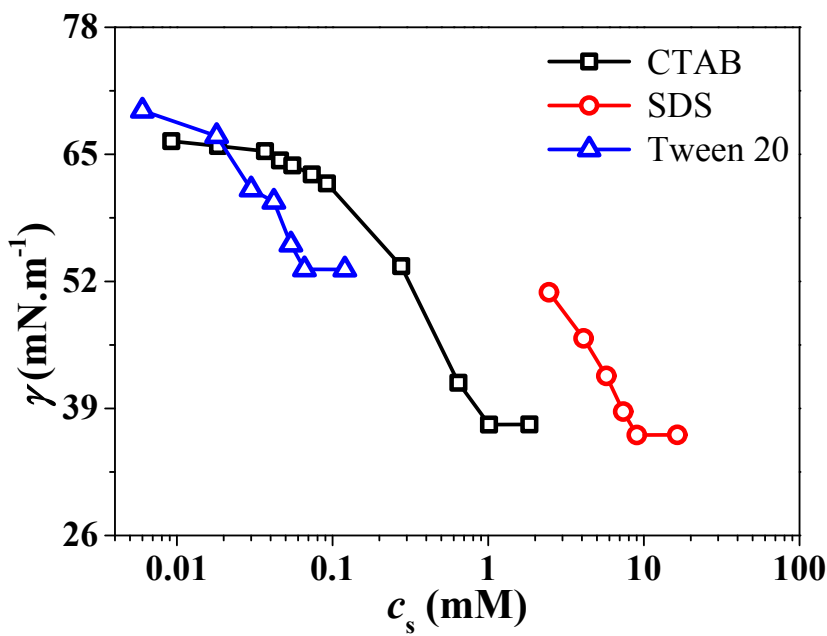

Figure 2. Surface tension in the presence of different surfactants.

## 3.3. Nanobubble Characterization

### 3.3.1. Nanobubble size and concentration measurement

The NTA (Malvern, NS300, UK) technique was used to determine the average bubble size and concentration of bulk NBs. The Nanosight NS300 can measure sizes ranging from 10 nm to 1000 nm in the concentration range of $10^7 - 10^9$ NBs/mL. A volume of approximately 1 mL of NBs was manually loaded into the O-Ring top-plate of the laser module to determine the concentration and size. The charge-coupled device (CCD) camera equipped with NTA records the videos of the Brownian motion of NBs at 30 frames per second (FPS). For each analysis, five videos of different portions of samples were recorded. The recording duration of each video was 60 s. Brownian motion of NBs was captured at camera level and screen gain of 12.0 and 3.0, respectively. The procedure was repeated three times for each experimental condition, using different samples from the same vials. The NTA determines the bubble size using the Stokes-Einstein equation, which is represented as

$$D = \frac{k_B T}{6\pi\mu r} \tag{18}$$

where $D$ is the diffusion coefficient and $\mu$ is the dynamic viscosity of the liquid. NTA performs a Brownian translational motion analysis on the NBs and determines the diffusion coefficients as $D = (\bar{x}, \bar{y})^2 / 4t$, where $(\bar{x}, \bar{y})^2$ refers to the mean squared displacement of the NBs in two dimensions, and $t$ is the time interval between successive displacement measurements.[63] Tracking analysis and evaluation of NBs were performed at a detection threshold of 25.

### 3.3.2. Zeta Potential measurement

The electrical potential of bubbles in bulk suspension is known as zeta potential $(\zeta)$. A nanoscale entity either possesses a +ive or −ive charge depending on the aqueous medium and attracts oppositely charged ions, forming an electric double layer at the bubble-liquid interface. The slipping plane is the outer edge of the electrical double layer that separates the NBs from the bulk solution. The number of ions and their valence in the slipping plane determine the magnitude of $\zeta$. The $\zeta$ of NBs in DI water were determined by Zetasizer (ZS90, Malvern, UK). Zetasizer calculates the $\zeta$ by estimating the electrophoretic mobility (the ratio of the drift velocity of the dispersed phase, $U$ to the strength of the applied electric field, $E$) of bubbles in the presence of an electric field to the electrode of opposite charge and then applying Henry's expression, which is based on the Smoluchowisk equation. Henry's relation is defined as $U_E = 2\Phi\zeta f(ka)/3\mu$, where $U_E$, $\Phi$, and $f(ka)$ are the electrophoretic mobility, dielectric constant of the medium, and Henry's function, respectively. Henry's function is 1.50 (Smoluchowski approximation) in aqueous colloidal suspension and 1 in nonaqueous media (Hückel approximation).[64] The $\zeta$ was calculated by setting the dielectric constant and refractive index of the liquid to 80 and 1.33, respectively. The refractive index of NBs was taken as 1.[65] For each experimental condition, several samples of NBs were prepared at different dilution factors and $\zeta$ were measured. The primary rationale for diluting the NBs concentrations was a relatively high NBs concentration per 1 mL of solution. In order to determine $\zeta$ of NBs in DI, the sample was diluted with NB-free DI water. Similarly, NBs in various pH and surfactant solutions were diluted with the same pH and surfactant solution that did not contain NBs. The $\zeta$ of each sample was measured 10 times; thus, the reported results are based on an average of 10 data points.

## 4. Results and Discussions

### 4.1. Generation of Bulk NBs

Before generating the NBs in DI water, the presence of nanoscale impurities in the form of NBs is tested by NTA. No impurities were observed. The typical image of NBs in different liquid mediums captured by NTA is shown in Figure 3.

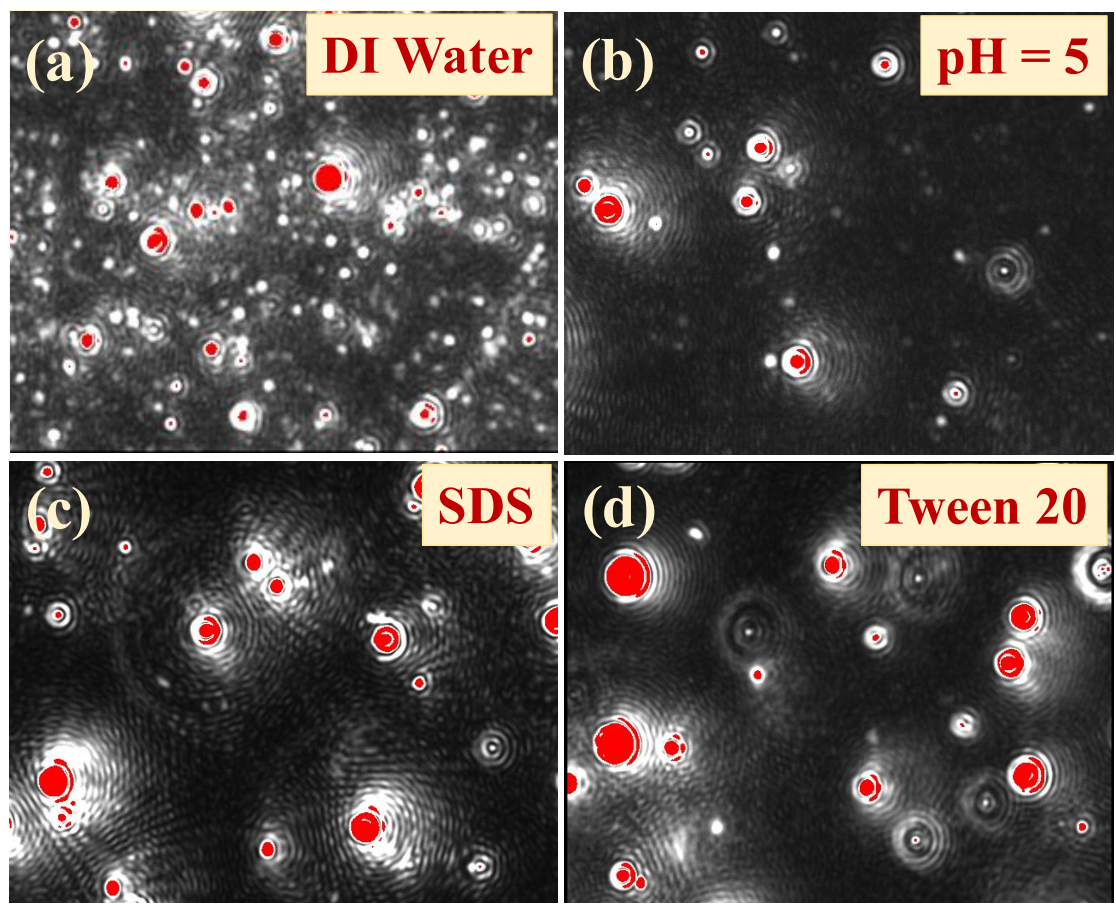

Figure 3. Typical images of NBs (at camera level of 12 and screen gain of 3) in various liquid solutions, captured by the NTA: (a) DI water, (b) pH solution (pH = 5), (c) SDS solution $(c_s = 4.1\,\text{mM})$, and (d) Tween 20 $(c_s = 3 \times 10^{-2}\,\text{mM})$. The larger NBs scatter more light (appearing red) and move more slowly than the smaller NBs, which move faster and scatter less light, appearing dimmer (looking white). Concentric rings surrounding the NBs represent the ripples that form in the medium owing to Brownian motion.

## 4.2. Zeta potential of NBs

The $\zeta$ of NBs influenced by different parameters such as pH, surfactant, electrolyte, temperature, density, and viscosity of the solution.[37,66] The adsorption of $H^+$ and $OH^-$ ions on the gas-liquid interface plays a decisive role in developing the electrical double layer around the bubble and surface charge.[67] When NBs are produced in a neutral liquid, an equal amount of $OH^-$ and $H_3O^+$ is produced. In an acidic solution, however, the production of $H_3O^+$ exceeds that of $OH^-$.[68] Short-lived $OH^-$ ions normally attach to the NBs surfaces and stabilize them; however, in an acidic solution, $H^+$ ions from the HCl take over and bind to the bubble surface, making it +ively charged.[69] The $\zeta$ of NBs in different pH media are presented in Figure 4a. At pH = 7, the $\zeta$ was − 25.41 mV. In the neutral condition, the slipping plane (outermost plane of EDL) of NBs has a substantial concentration of $OH^-$ ion; therefore, the $\zeta$ has a −ive value. Another study reported a −ive $\zeta$ of NBs in pure water.[70] It is observed that as pH reduces from 7 to 5, the magnitude of the −ive value of $\zeta$ reduces significantly. The $\zeta$ values of NBs in DI are −ive for pH > 4.5. At pH = 4.5, $\zeta$ of NBs is close to zero, reflecting an isoelectric point (IEP). For pH < 4.5, $\zeta$ values were +ive samely with Takahashi [71]. Calgaroto $et$ $al.$ [72] elucidated a sigmoidal behavior of the $\zeta$ curve in the pH range of 2 (+ 26 mV) to 8.5 (− 28 mV) with the isoelectric point at pH = 4.5. At pH = 10, they observed a maximum −ive $\zeta$ value of − 59 mV. The change in $\zeta$ from a −ive to a +ive value reflects preferential adsorption of the $H^+$ ion on the slipping plane. The magnitude of $\zeta$ indicates the longevity of NBs in the suspension. The longevity of NBs depends on the magnitude of surface charge, which governs the attractive van der Waals forces and the repulsive electrical forces. According to Spanos $et$ $al.$ [73], a stable aqueous dispersion is commonly defined as having a minimum absolute $\zeta$ value of 30 mV. The repulsive electrical forces developed by the charged surface of the bubble prevent the bubble from coalescence, consequently having a high $\zeta$ .[74] The magnitude of −ive $\zeta$ was observed to increase with an increase in pH from 7 to 11. This may be owing to preferential adsorption of $OH^-$ ion.

The surface charge of NBs in pure water is extremely pH-dependent. However, surfactant addition alters the surface characteristics at the gas-liquid interface, affecting the $\zeta$ of NBs.[71] The ionization of the surfactant mainly impacts surface characteristics at the gas-liquid interface. The critical micelle concentration (CMC) of CTAB is 0.92 mM. Surfactant molecules self-assemble into structured molecular assemblies called micelles in water.[75] The variation in $\zeta$ of NBs in the presence of different CTAB concentrations ($0.92 \times 10^{-2} - 184.0 \times 10^{-2}$ mM) is shown in Figure 4b. In the absence of CTAB, the $\zeta$ of NBs was − 25.42 mV. A small addition of CTAB, at $0.92 \times 10^{-2}$ mM, results in a significant reduction in the magnitude of the −ive value of $\zeta$ to − 7.15 mV. With a further increase in the CTAB concentration to $1.84 \times 10^{-2}$ mM, $\zeta$ changes from −ive to +ive values (+ 8.85 mV). In CTAB solution, the IEP of NBs is between $0.92 \times 10^{-2} - 1.84 \times 10^{-2}$ mM. The conversion of −ive to +ive $\zeta$ of NBs is due to the adsorption of +ive CTA$^+$ ions on the bubble surface. With increasing CTAB concentration, $\zeta$ increased and reached a maximum value of + 47.02 mV at $c_s = 64.4 \times 10^{-2}$ mM .

The influence of SDS surfactant at various concentrations on $\zeta$ is depicted in Figure 4c. The CMC of SDS is 8.2 mM. The $\zeta$ was detected over a wide range of SDS concentrations, both below and above the CMC. The $\zeta$ of NBs is −ive for all the SDS concentrations. The −ive magnitude of $\zeta$ increased from − 28.57 to − 51.63 mV as SDS concentrations are increased (from 2.46 to 7.38 mM) below the CMC threshold (8.2 mM). The magnitude of −ive $\zeta$ increases mainly because of the adsorption of $SO_4^-$ ions from the SDS. The increase in the −ive magnitude of $\zeta$ below the CMC of SDS is consistent with the findings of other investigators.[38,76] However, the −ive magnitude of $\zeta$ was found to be reduced above the CMC. At 16.4 mM (approximately double the CMC), the $\zeta$ decreases to − 39.48 mV.

The variation in $\zeta$ of NBs in the presence of a nonionic surfactant of Tween 20 is illustrated in Figure 4d. Tween 20 has a CMC of 0.06 mM. The influence of Tween 20 concentrations below and above the CMC is observed. As noticed, the −ive magnitude $\zeta$ of NBs in Tween 20 solution is lower for all the concentrations compared to the DI water. Tween 20 does not contain any charge; thus, its presence has little influence on the $\zeta$ of NBs. With increasing Tween 20 concentration (below CMC), $\zeta$ varies in the range of − 13.25 mV to − 21.30 mV. Above the CMC ($c_s = 6.60 \times 10^{-2}$ mM), no significant variation in $\zeta$ was observed. It can be concluded that in the presence of Tween 20, no discernible trend in NBs $\zeta$ was noticed.

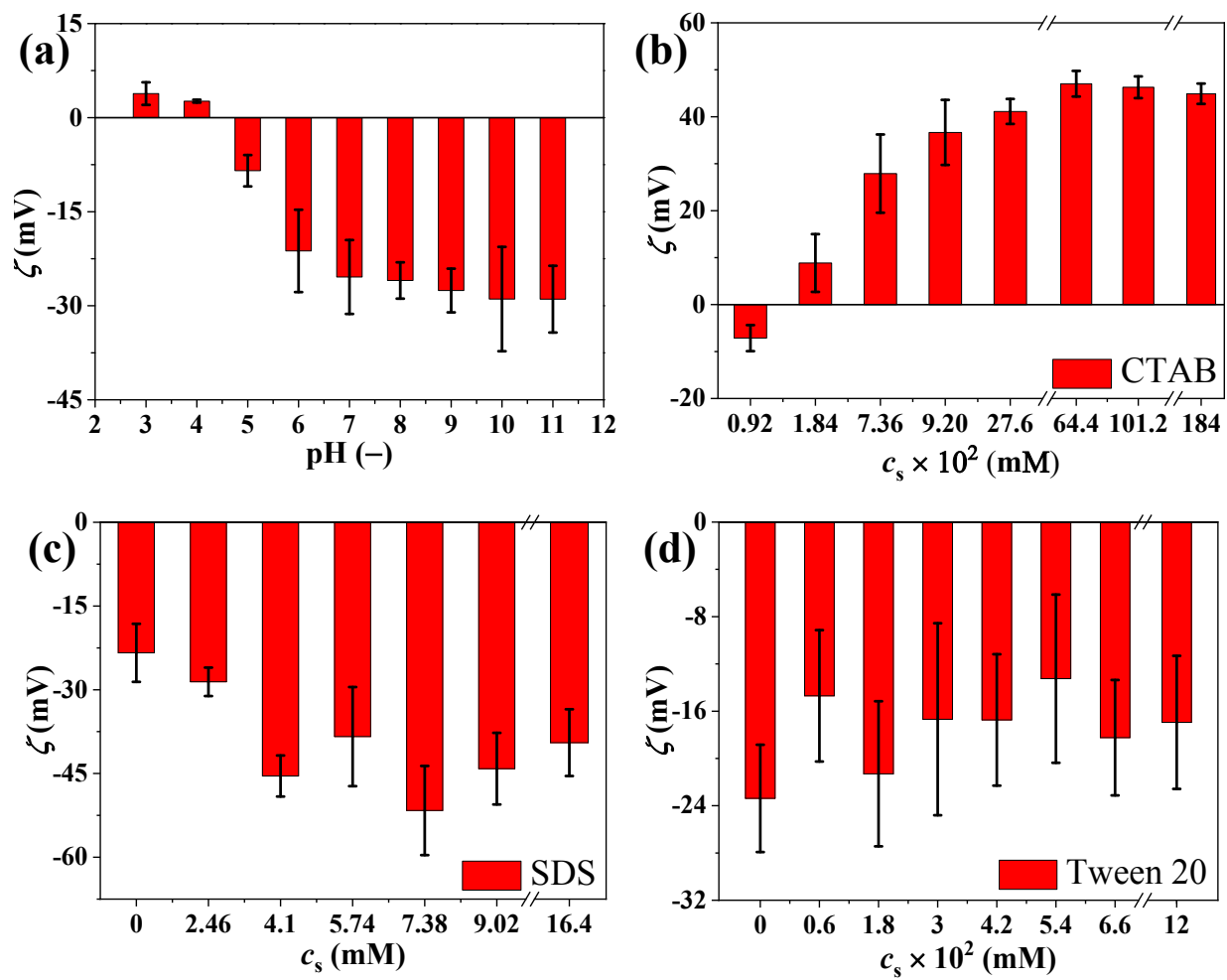

Figure 4. Zeta potential of NBs in different solutions: (a) pH, (b) CTAB, (c) SDS, and (d) Tween 20.

## 4.3. Variation in NBs size in different solutions

### 4.3.1. Effect of pH on NBs size

The effect of pH on NBs size is shown in Figure 5a. The average bubble size in an acidic medium (pH < 7) is higher compared to the neutral medium (pH = 7). The maximum value of the average bubble size is 215.70 nm at IEP (pH = 4.5), which is 87% higher compared to the neutral medium. The ion stabilization theory can explain the larger bubble size close to IEP. Charged NBs gives rise to an electrostatic pressure surrounding the bubble.[24] The electrostatic pressure balances the internal Laplace pressure; as a result, diffusion of gases to the environment can be prevented.[77] The accumulated ions around the bubble surface form a thin coating that serves as a diffusion barrier, limiting gas dissolution and thus increasing the longevity of the NBs, a process known as the ion shielding effect.[78] At equilibrium, internal Laplace pressure and the electrostatic pressure provide a relationship between the NBs radius and surface charge density, as reported by Bunkin and Bunkin [24]

$$r = \frac{\gamma \varepsilon}{\pi \sigma_0^2} \tag{19}$$

The inverse relationship between bubble radius and surface charge density, as depicted in Eq. (19), indicates that the smaller the surface charge density, the larger the bubble size, as evident in Figure 5a. Above IEP, the surface charge of NBs is −ive, although its magnitude decreases with a reduction in pH. The reduction in −ive pH magnitude leads to an increase in the average bubble size, as can be seen at pH = 6 and pH = 5. Below IEP, the bubble surface carries a +ive charge, although the magnitude of the charge is not as high as evident in Figure 4a. As pH reduces from 4 to 3, the average bubble size reduces from 181.40 nm to 157.50 nm. Results show that the higher the magnitude of the charge of the NBs, the smaller the average bubble size. In both −ively and +ively charged NBs, the smallest bubble size corresponds to the largest magnitude of charge.

### 4.3.2. Effect of CTAB on NBs size

The influence of CTAB on NBs average bubble size is demonstrated in Figure 5b. As noticed, NBs size is larger for all the CTAB concentrations compared to DI water. This may be attributed to the variation in surface charge of NBs in the presence of CTAB. It is well known that the addition of surfactant reduces the surface tension of the liquid and consequently stabilizes and reduces the millimeter-sized bubbles, as evident in the literature.[79] For these systems, bubble size and bubble density are considerably dependent on the physical properties of the medium, such as surface tension, viscosity, and density of the liquid. However, in the case of NBs, the size depends on the magnitude of absolute surface charge rather than the surface tension of the medium, as evident in Figure 5b. It can be noticed that when a +ive charge is established on the NBs surface at $1.84 \times 10^{-2}$ mM, the further addition of CTAB results in two important effects: (i) enhancement in the magnitude of the +ive charge of NBs and (ii) a reduction in the surface tension of the medium. In this scenario, the former has a more substantial influence than the latter. Bubble diameter at the initial concentration of CTAB ($c_s = 0.92 \times 10^{-2}$ mM) is 2.94 times higher than DI water. It may be due to the reduction in the magnitude of the −ive charge from − 25.42 to − 7.15 mV. The reduction in the magnitude of charge prefers the attractive van der Waals force between the bubbles rather than the repulsive force. The higher magnitude of the attractive van der Waals force results in the bubble coalescence. As a result, the bubble size grew. The average bubble size is observed to decrease as the surfactant concentration increases above $1.84 \times 10^{-2}$ mM. The minimum bubble size in CTAB is observed to be 160.2 nm at $184 \times 10^{-2}$ mM, which is 1.39 times the NBs in the absence of CTAB.

### 4.3.3. Influence of SDS on NBs size

The size of the NBs in DI water is smaller than in the SDS system, as depicted in Figure 5c. The NBs size at 2.46 mM (a minimum SDS concentration) is 1.31 times the size in DI water. The bubble size slightly decreases from 150.65 nm to 141.75 nm as the SDS concentration increases from 2.46 to 7.38 mM, approaching close to the CMC. It can be demonstrated that at high concentrations of SDS, close to the CMC, the reduction in the size of NBs is only 5.91%, which is insignificant. The reduction in NBs size with an increase in the SDS concentration is consistent with the results of another investigator,[38] where they varied the SDS concentration from 1 to 8.1 mM, which was below the CMC. Phan *et al*.[76] also demarcated a reducing bubble size by increasing the SDS concentration. By increasing the SDS concentration over the CMC, the size grows to 152.75 nm (at 9.02 mM, 1.1 times the CMC) and 168.15 nm (at 16.4 mM, 2 times the CMC). Over CMC of SDS, the magnitude of $\zeta$ of NBs also decreases to $-39.77$ mV at 16.4 mM, compared to $-51.63$ mV at 7.38 mM, as evident in Figure 4c. The reduction in the magnitude of $\zeta$ destabilizes the NBs size preferentially by reducing repulsive force between the bubbles. Consequently, bubble size increases due to coalescence. Above the CMC, the surfactant starts to form micelles rather than effectively reduce the liquid cohesive force between the water molecules and may adsorb to the NBs surface, forming an additional layer. The formation of an additional layer may also increase the NBs size. Researchers have reported the SDS micelle size in the range of $1.60 - 4.0$ nm,[80-82] which is too small to be detected by an NTA.[36] The primary interactions that occur during micelle formation are electrostatic interaction, van der Waals forces, hydrogen bonding, and hydrophobic effects.[60] At CMC, 50% of the surfactant is in micelles, whereas 50% is a free surfactant. Increases in the surfactant concentration result in a more significant percentage of the surfactant being incorporated into micelles.

### 4.3.4. NBs size in the presence of Tween 20

The variation in NBs size in the presence of Tween 20 surfactant is demonstrated in Figure 5d. It can be noticed that the NBs size is larger in all Tween 20 solutions compared to the DI water. No particular trend in NBs size as a function of Tween 20 below the CMC $(c_s \leq 5.40 \times 10^{-2}\,\mathrm{mM})$ was observed. However, there is a significant increase in bubble size above the CMC $(c_s \geq 6.60 \times 10^{-2}\,\mathrm{mM})$. At minimum surfactant concentration $(c_s = 0.60 \times 10^{-2}\,\mathrm{mM})$, the average bubble size is 127.90 nm, while the bubble size is 176.47 nm at $c_s = 12.0 \times 10^{-2}\,\mathrm{mM}$ (2 times the CMC). The bubble size at 2 CMC is 1.53 and 1.38 times larger than DI water and the solution with the minimum surfactant concentration, respectively. Basheva $et\ al.$ [83] investigated the size distribution of the micelles of Tween 20 in DI water at a 1.0 mM (about 20 CMC) concentration. At this condition, the average diameter of the micelle was approximately 7.2 nm. The maximum value of Tween 20 concentration in this study is 2.0 CMC, which is significantly lower than the concentration reported by Basheva $et\ al.$ [83]. As a result, the micelle size in the current experimental condition will be much smaller, which the NTA cannot detect and thus cannot influence the result.

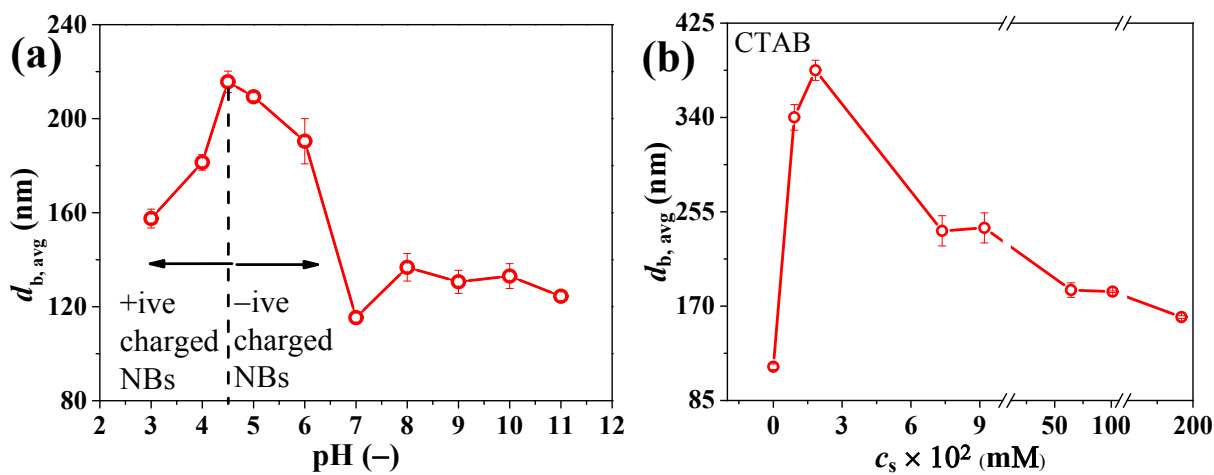

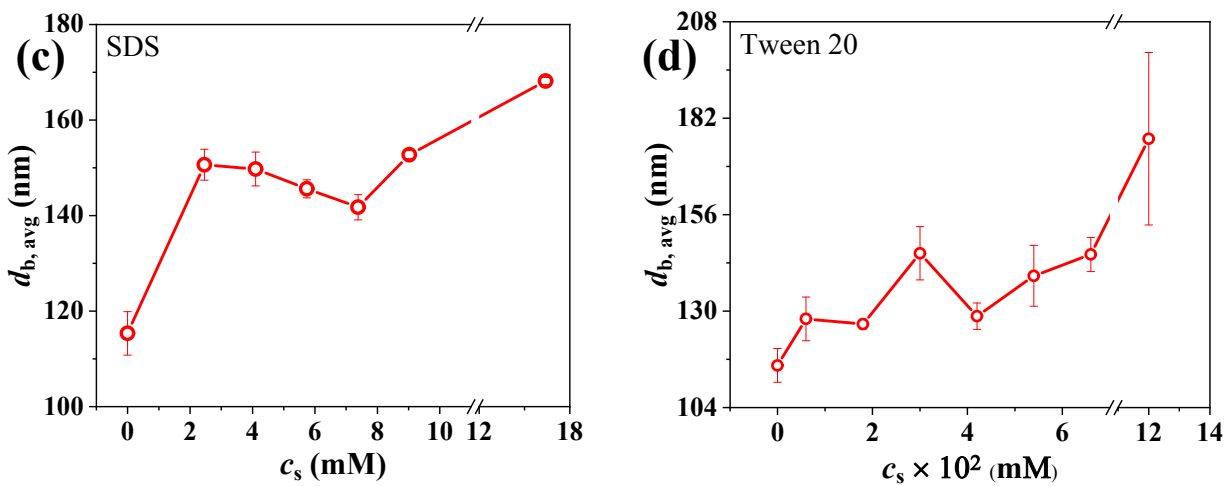

Figure 5. Variation in NBs average bubble size in the presence of (a) pH, (b) CTAB, (c) SDS, and (d) Tween 20.

### 4.4. Bubble size distribution and corresponding concentrations

The variation in bubble size distribution (BSD) and NBs concentration ($c_{nb}$) in different experimental conditions are shown in Figures 6a to 6h. The effect of pH on NBs BSD is shown in Figure 6a. The x-axis refers to the bubble size, while the y-axis refers to the relative frequency (RF) of bubbles. As observed, BSD at pH = 7 is aligned to the smaller bubble size compared to the other pH conditions. BSD at pH = 5 (close to IEP) was observed to be at a larger bubble size than other pH conditions, which is attributed to instability in the surface charge of the bubble that leads to bubble coalescence. BSD curves shift to the smaller bubble size as pH reduces from 6 to 4 and to 3. In an alkaline medium (pH = 10), the BSD curve is just the right-hand side of the pH = 7, attributing a smaller NBs than in the acidic medium. BSD results are in agreement with the NBs average size, as reported in Figure 5a. The maximum RFs are observed to be at 102.5 nm (at pH = 3), 153.5 nm (pH = 4), 347.5 nm (pH = 5), 193.5 nm (pH = 6), 65.5 nm (pH = 7), and 111.5 nm (pH = 10). Shifting the maximum RF to the large bubble size indicates an increase in the average bubble size. In the case of pH = 5, 50% of the bubbles are below 324.5 nm, whereas in pH = 7, 50% of the bubble are below 73.5 nm. The percentage of the bubble is calculated from the cumulative BSD. In the case of pH = 6, 4, and 3, 50% of the bubbles are below 226.5 nm, 162.5 nm, and 155.5 nm, respectively. The $c_{nb}$ analysis in various pH media is demonstrated in Figure 6b. The $c_{nb}$ at different pH levels are listed in the following order: pH = 7 ($1.94 \times 10^9$ NBs/mL) >

pH = 10 (2.15 × 10$^8$ NBs/mL) > pH = 6 (1.41 × 10$^8$ NBs/mL) > pH = 3 (1.33 × 10$^8$ NBs/mL) > pH = 4 (1.32 × 10$^8$ NBs/mL) > pH = 5 (1.20 × 10$^8$ NBs/mL). The $c_{nb}$ in pH medium is low compared to the DI water. At pH = 5, the $c_{nb}$ is at its lowest value.

BSD at various concentrations of CTAB is shown in Figure 6c. The BSD curve at $c_s = 0.92 \times 10^{-2}$ mM concentration is aligned toward the larger bubble size than other concentrations. The BSD curve shifted to the slightly smaller bubble size with a further increase in the CTAB concentration at 1.84 × 10$^{-2}$ mM. Two high peaks can be observed at this concentration, indicating higher fluctuations in BSD. At this concentration, NBs $\zeta$ change from −ive to +ive, as evident from Figure 4b, causing instability in attractive van der Waals and repulsive forces. BSD curves shift to the small bubble size as CTAB concentrations exceed 3.68 × 10$^{-2}$ mM. The maximum RF is at $c_s = 64.4 \times 10^{-2}$ mM , which is close to the CMC. BSD result agrees with the results of average bubble size as demarcated in Figure 5b. The $c_{nb}$ in CTAB solutions are depicted in Figure 6d. The $c_{nb}$ increases from 7.11 × 10$^8$ NBs/mL to 1.55 × 10$^9$ NBs/mL as the CTAB concentration increases from 0.92 × 10$^{-2}$ mM to 7.36 × 10$^{-2}$ mM. The further increase in the CTAB concentration approaching the CMC reduces the $c_{nb}$. At $c_s = 184 \times 10^{-2}$ mM (2 times the CMC), the $c_{nb}$ is 9.14 × 10$^7$ NBs/mL. The significant reduction in $c_{nb}$ is attributed to the micelle formation.

Figure 6e depicts the variation in BSD at various SDS concentrations. No significant variation in BSD curves was observed in the presence of SDS. It is because of no change in the $\zeta$ of NBs from −ive to +ive or vice versa (as evident from Figure 4c). In the presence of SDS, NBs carry a −ive charge, and the magnitude of the −ive charge increases with an increase in the SDS concentration up to the CMC. The maximum RF peaks vary from 129.5 nm to 146.5 nm as the SDS concentration increases from 2.46 to 16.4 mM. Below the CMC ($c_s \leq 7.38$ mM) , the 50% bubble of total NBs, lies in the range of $132.5 \, \text{nm} \leq d_b \leq 151.5 \, \text{nm}$ , while above the CMC, this range is $154.5 \, \text{nm} \leq d_b \leq 161.5 \, \text{nm}$ . The existence of NBs at a larger bubble size range leads to an increase in the average bubble size above the CMC. The BSD result is concurrent with the results of average bubble size, as exhibited in Figure 5c. The $c_{nb}$ in SDS solutions were quantified, as shown in Figure 6f. It is observed that the $c_{nb}$ increases with an increase in the SDS concentration below the CMC; however, above the CMC, the concentration reduces significantly. The $c_{nb}$ result in the presence of SDS (concentration up to CMC) is consistent with the result of another researcher.[38] The observed $c_{nb}$ ranges from $2.94 \times 10^8 - 3.38 \times 10^8$ NBs/mL (in the $c_s = 2.46 - 7.38$ mM ). The increase in the $c_{nb}$ is about 14.97% when the surfactant concentration increases from 2.46 to 7.38 mM. At 9.02 and 16.4 mM (above the CMC), $c_{nb}$ reduces to $3.15 \times 10^8$ and $2.76 \times 10^8$ NBs/mL, respectively. By comparing the $c_{nb}$ at the maximum value of SDS concentration (below and above the CMC), the reduction in $c_{nb}$ is noticed to be around 18.34%. In conclusion, it can be said that the increase in the SDS concentration up to the CMC leads to an increase in the $c_{nb}$, whereas above the CMC, it reduces. The $c_{nb}$ has a substantial impact on the average bubble size of NBs.

BSD in the presence of Tween 20 is illustrated in Figure 6g. No particular trend in the BSD curve is observed. Several RF peaks can be observed for each BSD curve. As noticed in Figure 4d, the $\zeta$ of NBs do not have any regular trend; however, all values are –ive. The magnitude of the –ive value fluctuated, hence the several peaks in BSD. The maximum RF of curves varies in the range of 62.5 nm to 89.5 nm in the Tween concentration range of $0.6 \times 10^{-2}$ to $5.4 \times 10^{-2}$ mM (below the CMC). Above the CMC of Tween 20, the maximum RF shifted to 121.5 nm and 238.5 nm at $6.6 \times 10^{-2}$ and $12 \times 10^{-2}$ mM, respectively. As manifested in Figure 5d, shifting BSD to the larger bubble size indicates an increase in the average bubble size. The $c_{nb}$ is also analyzed, as illustrated in Figure 6h. The $c_{nb}$ was observed to vary in the range of $1.49 \times 10^8 - 1.50 \times 10^8$ NBs/mL, with a Tween 20 concentration range of $0.6 \times 10^{-2}$ to $5.4 \times 10^{-2}$ mM (below the CMC). Above the CMC, $c_{nb}$ falls to $1.40 \times 10^8$ NBs/mL and $1.07 \times 10^8$ NBs/mL at $6.60 \times 10^{-2}$ mM and $12 \times 10^{-2}$ mM, respectively.

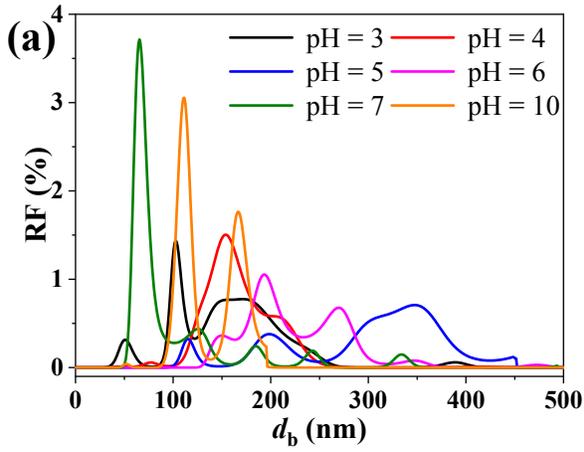
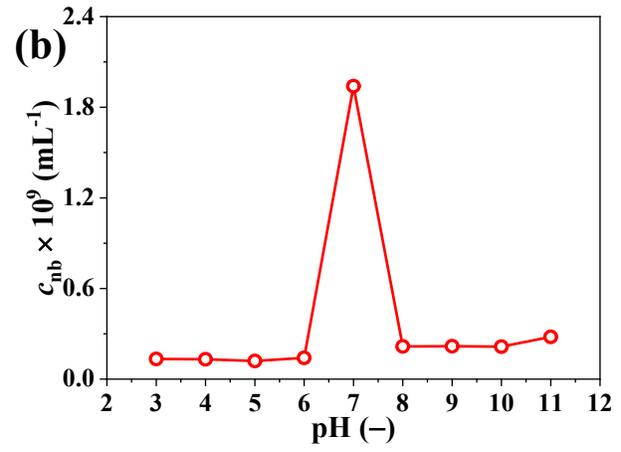
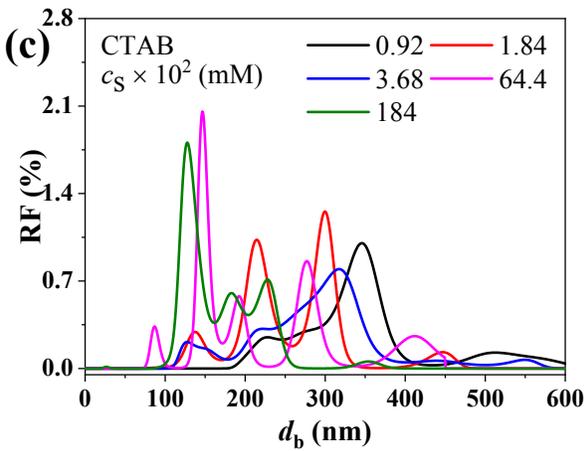
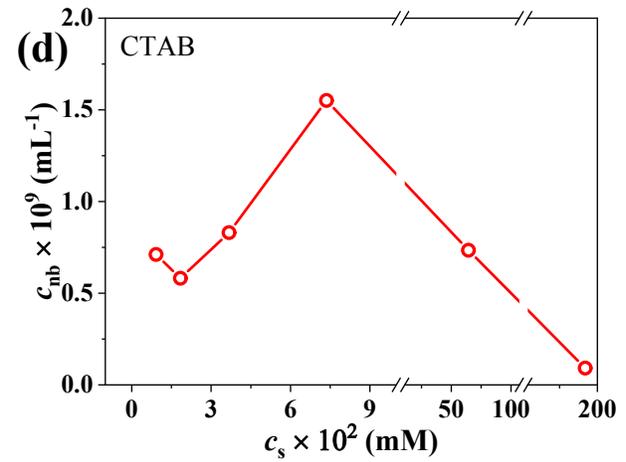
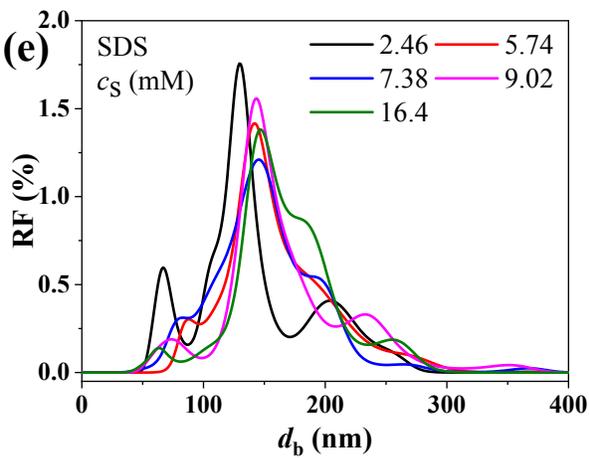
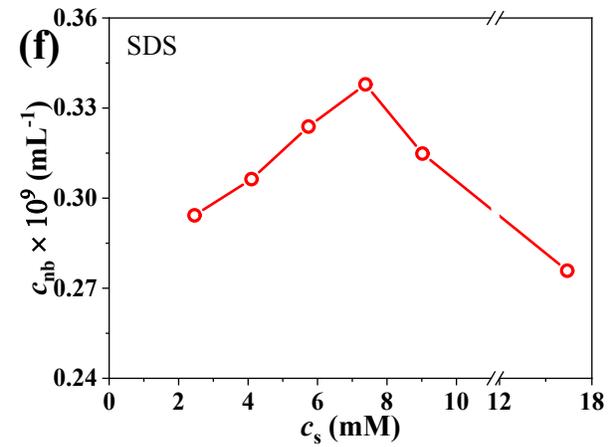

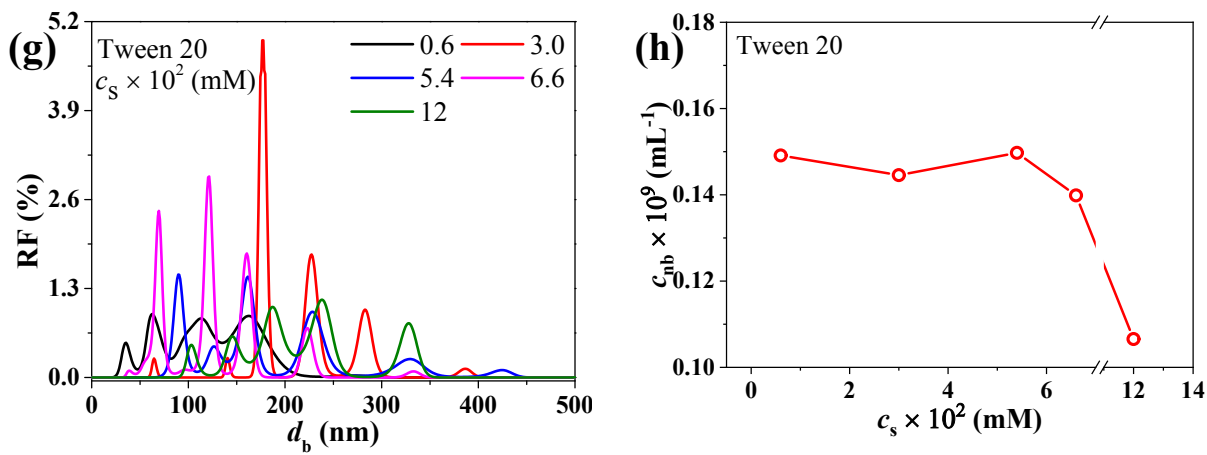

Figure 6. Relative frequency and NBs concentration in various experimental conditions: (a) and (b) pH medium, (c) and (d) CTAB solutions, (e) and (f) SDS solutions, and (g) and (h) Tween 20 solutions.

## 4.5. Stability of NBs

The most remarkable property of NBs is their incredible longevity; they have been observed to persist for days, weeks, or even months.[18,21,22] In this section, the stability of NBs over time in terms of average bubble size and concentrations is shown in Figures 7a to 7h. The variation in bubble size and concentration in various experimental conditions is monitored for different periods of time. The changes in average size over time in different pH solutions are illustrated in Figure 7a. As observed, the average size continues to grow over time at pH = 7. If compared from day 1 ($d_{b,avg} = 115.35 \pm 2.55$ nm), significant growth in bubble size can be observed on day 11 ($d_{b,avg} = 206.27 \pm 3.76$ nm). The $d_{b,avg}$ on day 11 is about 1.79 times that of day 1. This result can be explained by the fact that bubble nucleation is high as long as there is enough dissolved gas to provide the required nuclei. After day 11, the growth in bubble size is minimal and reaches a value of 200 nm on day 52. NBs at pH = 7 is mainly stabilized by their significant surface charge, resulting from the adsorption of hydroxyl ions created by the self-ionization of water. The behavior of growth and reduction in bubble size over time in DI water can be conjectured by either the Ostwald ripening or the Smoluchowski ripening effect.[22,84] Big bubbles expand while small bubbles contract during Ostwald ripening. In the liquid phase, gas molecules that have detached from small bubbles join big NBs. This coarsening process occurs at the nanoscale level. Another investigator also discerns the similar characteristics of NBs over time in DI water and postulates that the Ostwald ripening process is accountable for them.[22] Lee *et al*. [84] noticed the continuous growth in the NBs size from day 1 to day 6 and speculated that the Ostwald ripening is the reason for such behavior. Smoluchowski ripening, on the other hand, is triggered by collisions between NBs as they move in the liquid phase. The Brownian motion of the bubbles mostly causes collisions between NBs. In this research, Smoluchowski ripening may be the reason for the NBs behavior over time.

At pH = 5, a linear rise in bubble size can be noticed up to day 18; beyond that, a significant bubble size reduction was observed. A linear rise in the bubble size occurs because the liquid medium is close to the IEP of pH = 4.5, which reduces the magnitude of the repulsive force between the bubbles. As a result, bubble coalescence occurs, resulting in a larger bubble size. Reduction in bubble size indicates collapsing of larger bubbles over time because of the imbalance in internal Laplace pressure and electrostatic pressure of the bubble. In general, bubbles vanish as (i) if they dissolve, (ii) bubble collision and coalescence, and (iii) rising of the bubbles to the liquid surface because of buoyancy force.[85] A bubble smaller than 1 $\mu$m in diameter may have an extremely low terminal velocity (~ 1 $\mu$m/s), preventing it from floating due to the buoyancy force.[22] Another study found that the bubbles with a size less than 5 $\mu$m have very little buoyancy and do not ascend to the surface.[86] Buoyancy cannot be ignored over time in this work, even though bubble sizes are below 1 $\mu$m. As can be seen that the reduction in bubble size is significant for a few liquid mediums over time. It may be possible that a larger bubble may reach the liquid surface and dissipate over time. Over time, a reduction in bubble size is observed for all the pHs. No NBs were observed beyond day 18 at a low pH = 3, indicating that the liquid medium was unsuitable for long-term stability.

For all basic media of pH = 8 to 11, a considerable rise in the average bubble size of $d_{b, avg} = 124.45$ nm to 193.0 nm was observed from day 1 to day 11, after which the average size witnessed a minor increase and remained stable at 195.80 nm to 205.40 nm for 52 days. A considerable variation in bubble size was seen in the acidic media on all the days but not in the basic medium. This reflects the fact that NBs are more stable in basic media than acidic media.

The behavior of the $c_{nb}$ over time at different pHs is demonstrated in Figure 7b. At pH = 5, 6, and 7 over time, $c_{nb}$ was observed to increase from $1.20 \times 10^8$ to $2.14 \times 10^9$ NBs/mL, while at pH =

3, and 4, it decreased from $1.33 \times 10^8$ to $2.01 \times 10^7$ NBs/mL. The dissipation of large bubbles and the availability of a large number of tiny bubbles may explain the rise in $c_{nb}$ over time. When the bubble size in a certain volume of NBs suspension is lowered, the concentration of the bubbles rises. In the basic medium (pH $= 8 - 11$), $c_{nb}$ diminished with time for all pH values. Over 52 days, at pH $= 8$, $c_{nb}$ reduces from $2.17 \times 10^8$ to $1.66 \times 10^8$ NBs/mL, a drop of about 28.38%. Similarly, at pH $= 11$, $c_{nb}$ decreases from $2.80 \times 10^8$ to $1.88 \times 10^8$ NBs/mL, a depletion of approximately 32.74%.

NBs average bubble size in different CTAB solutions was observed for 56 days, as shown in Figure 7c. For concentrations below the CMC of $64.4 \times 10^{-2}$ mM, a considerable variation in average bubble size was observed for all the days. Bubble size was seen to increase in some CTAB concentrations while decreasing in others. The maximum increase in the bubble size is 26.27% at $c_s = 7.36 \times 10^{-2}$ mM, with the bubble size increasing from 237.87 nm to 300.35 nm. The maximum reduction in the bubble size is 16.95%, with the bubble size decreasing from 382.50 nm to 317.65 nm at $c_s = 1.84 \times 10^{-2}$ mM. A slight increase in bubble size above the CMC was noticed in 12 days and remains stable in the range of 170.55 to 175.87 nm on day 56. The variation in NBs size above the CMC is minimal. Figure 7d depicts the concentrations of NBs in varied CTAB concentrations over time. On 56 days, a substantial decline in $c_{nb}$ was found for all CTAB concentrations.

Figure 7e depicts the NBs size in the SDS solution, which ranges from 141.75 to 168.15 nm on day 1. In 35 days, the bubble size increased steadily in the range of 156.65 to 203.90 nm. After 35 days, bubble size slightly decreased and was stable between 155.20 and 182.45 nm for 55 days. Figure 7f exhibits the $c_{nb}$ in different SDS concentrations. The concentration of NBs in some SDS solutions diminishes with time, whereas it increases in others. On day 52, $c_{nb}$ were in the range of $1.26 \times 10^8$ to $4.90 \times 10^8$ NBs/mL. Figures 7g and 7h depict the NBs size and concentration in Tween 20 solutions. Over the course of 54 days, the average bubble size in several Tween 20 solutions decreased while the concentration increased.

The existence of NBs over a long time reflects the low impact of bubble coalescence and Smoluchowski ripening in pH and surfactant solutions. In some cases, the bubble coalescence and Smoluchowski ripening may be active significantly. Hence, the larger bubble size and low $c_{nb}$ were observed. The persistence of NBs surface charge and its magnitude over time govern the bubble coalescence, hence the stability.

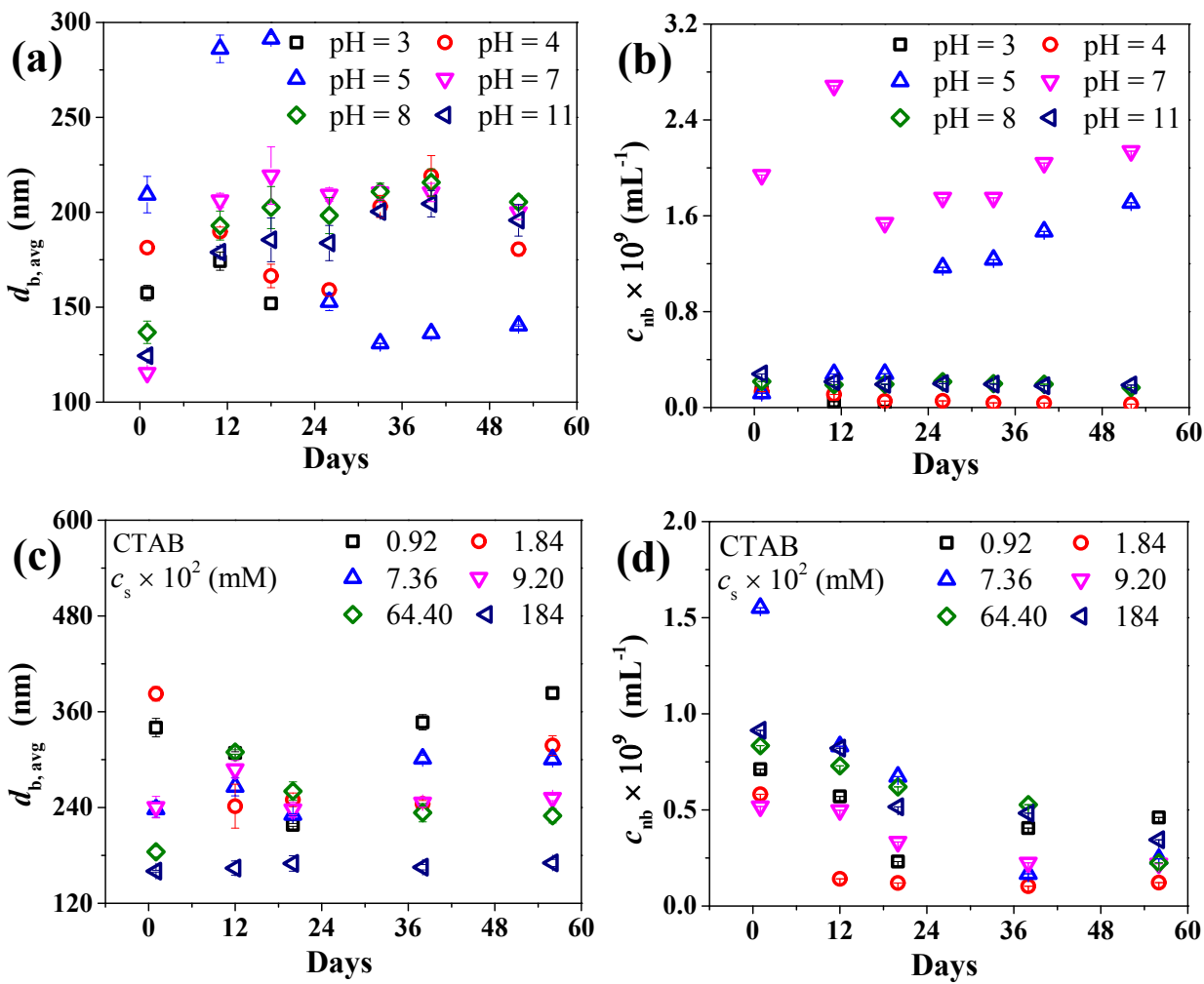

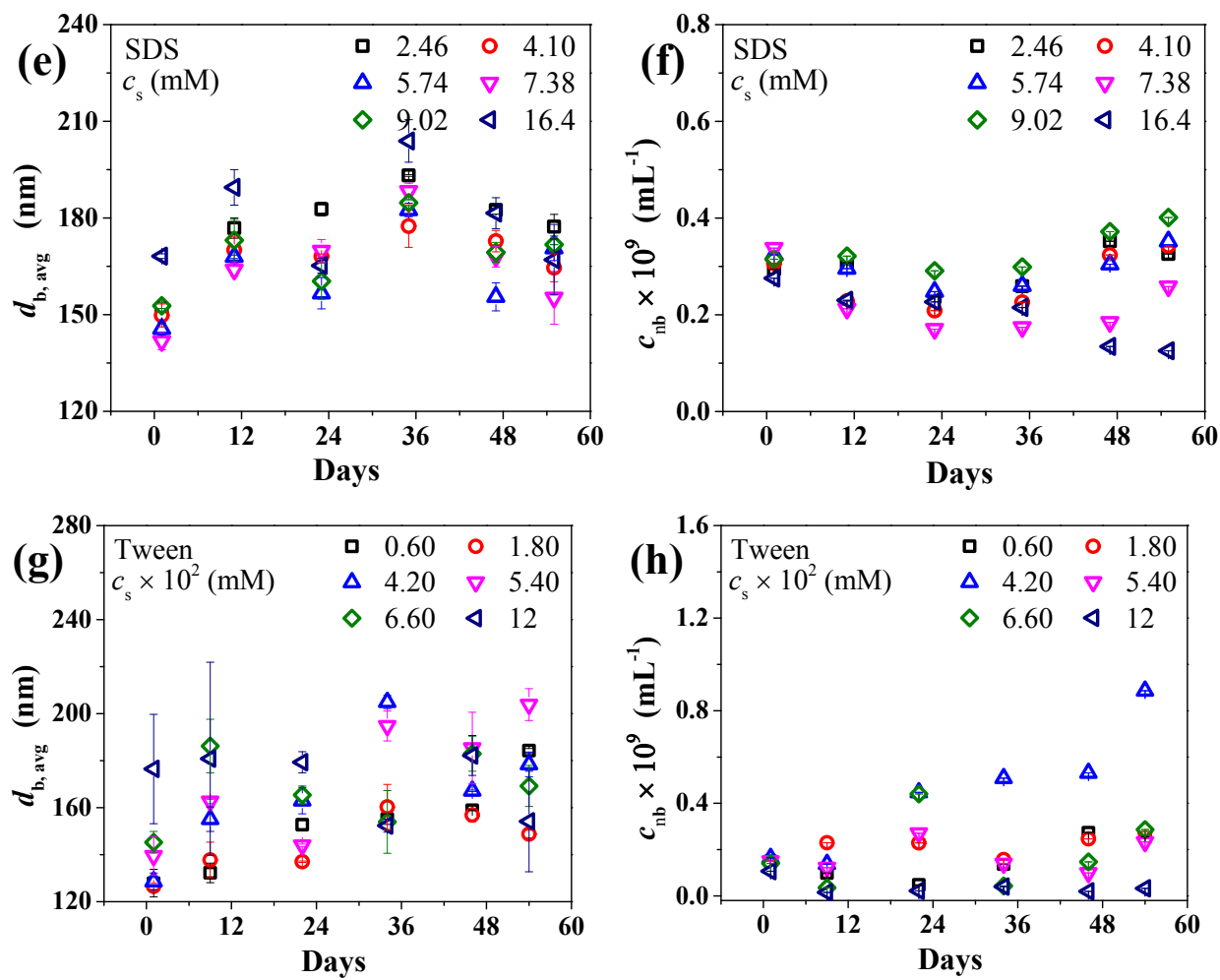

Figure 7. Evolution of NBs average bubble size and concentration over time: (a) $d_{b,avg}$ in pH, (b) concentration in pH, (c) $d_{b,avg}$ in CTAB, (d) NBs concentration in CTAB, (e) $d_{b,avg}$ in SDS, (f) concentration in SDS, (g) $d_{b,avg}$ in Tween 20, and (h) concentration in Tween 20.

## 4.6. Interpretation of NBs Stability

Interaction between NBs based on DLVO theory is depicted in Figures 8a – 8c. The potential barrier of normalized total interaction potential as a function of the separation distance ($d$) between two NBs can be used to determine stability. The higher the potential energy barrier, the lower the probability of bubbles coming close to coalescence. The total interaction potential for different pHs is depicted in Figure 8a. For $6 \leq pH \leq 11$, the potential energy barrier is +ive and greater than $43.90 k_B T$, whereas for pH < 6, the potential energy barrier almost vanishes. A low value of the potential energy barrier indicates van der Waals interaction potential dominance compared to electrostatic repulsive interaction potential. At higher potential energy barriers, NBs are more stable. A total potential energy barrier of $\sim 15 k_{\mathrm{B}} T$ or greater is required for stable dispersion,[87] while others have suggested $\sim 20 k_{\mathrm{B}} T$.[88] Another investigator also delineated a potential energy barrier of $20 k_{\mathrm{B}} T$ to $60 k_{\mathrm{B}} T$ in the range of pH = 4 to 10.[36] In another study, the maximum value of the total potential energy barrier for the NBs composed of different gas compositions was $30 k_{\mathrm{B}} T$.[22]

To get an in-depth insight into stability, the attractive van der Waals interaction potential as a function of separation distance at different pH mediums is shown in Figure 8b. All values are −ive in sign, reflecting the attractive nature of van der Waals interaction potential.[89] The high value is related to pH = 5, reflecting a substantial attractive force between the NBs, as pH = 5 is close to the IEP. Results with pH = 5 are in agreement with Figure 4a and Figure 5a. As evident in Figures 4a and 5a, bubble size was larger due to bubble coalescence as a result of dominant attractive van der Waals interaction potential. Zeta potential of NBs close to IEP leads to having the prevailing attractive van der Waals interaction potential. Moreover, pH = 3, and pH = 4 also have a considerable attractive van der Waals interaction potential. A slight variation in van der Waals interaction potential can be observed for $5 < \mathrm{pH} \leq 11$. It is also observed that the magnitude of van der Waals interaction potential decreases as the separation distance between NBs increases. It indicates that the van der Waals interaction potential is high when NBs are close.

A typical result of all the interaction potentials for a particular experimental condition (at pH = 6) is illustrated in Figure 8c. As can be seen, the electrostatic repulsive interaction potential ($E_{\mathrm{R}}$) decreases while the hydrophobic interaction potential ($E_{\mathrm{H}}$) increases with increasing distance between NBs. The values of attractive van der Waals interaction potential are close to zero. Total interaction potential is shifting from −ive to +ive with an increase in the separation distance. The maximum value of the total interaction potential is observed to be $43.90 k_{\mathrm{B}} T$ (at a 100 nm separation distance). The trend of these forces between two NBs is consistent with the result of another researcher.[22]

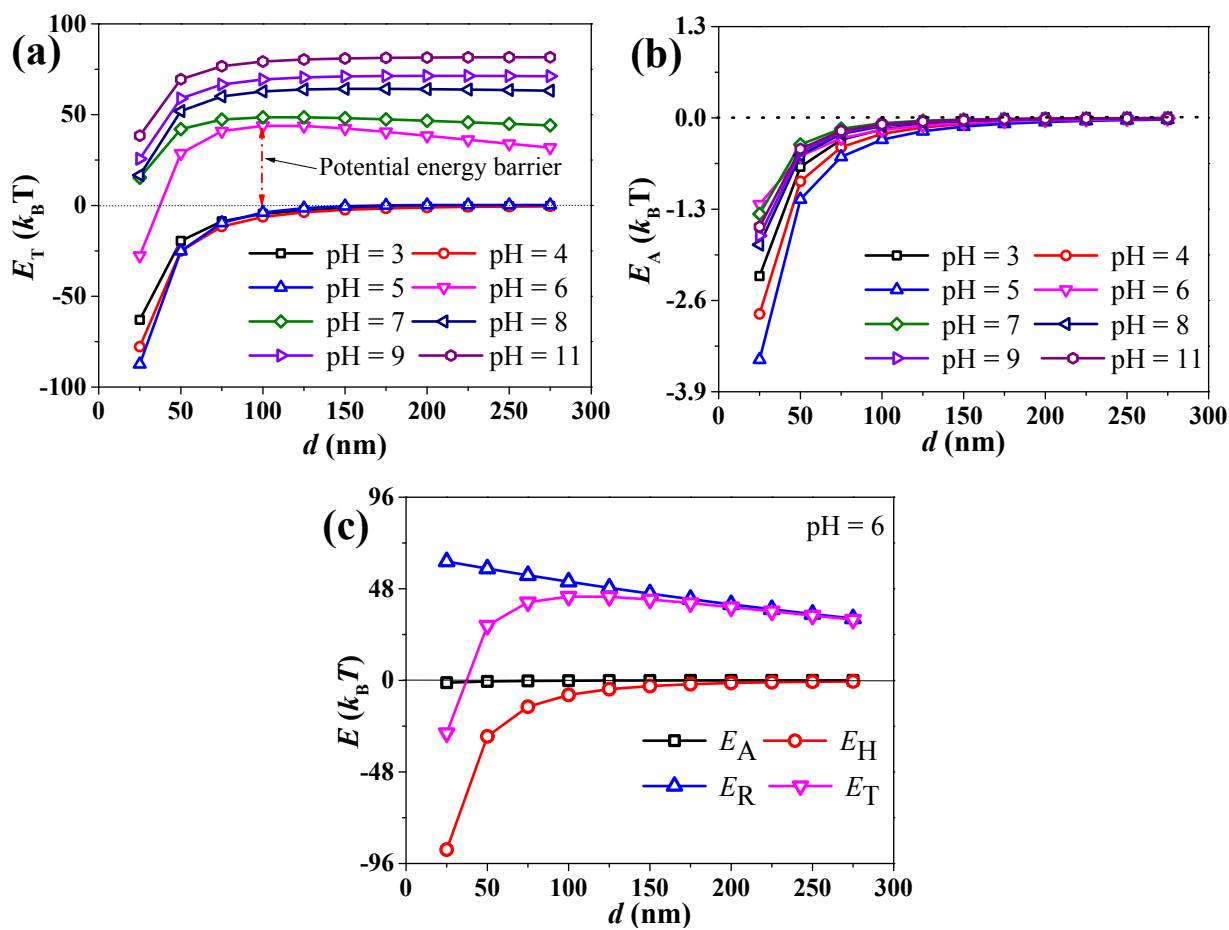

Figure 8. Interaction potential between two NB: (a) total interaction potential for different pH media, (b) van der Waals interaction potential, and (c) typical DLVO interaction containing all potentials ($E$ on the y-axis refers to the DLVO interaction potential).

## 4.7. Adsorption of surfactants at NBs surface

The calculated Debye lengths for SDS and CTAB solutions using Eq. (17) are in the range of 0.53 – 3.21 nm and 0.74 – 11.22 nm, respectively. In the presence of Gemini surfactant, Phan *et al.* [62] observed a Debye length of 6.2 Å on the gas-liquid interface rather than on the NBs surface. The Debye lengths were in the same order as the hydrated anions. Another researcher obtains the Debye length around silica in the presence of SDS by fitting the force curve to the DLVO theory.[90] Debye lengths ranged from 4.32 to 9.63 nm. CTAB and SDS solutions exhibited molecular areas (determined using Eq. 16) of surfactant at the interface ranging from 0.58 to 38.30 nm²/molecule and 0.57 to 13.67 nm²/molecule, respectively.

NBs colloidal properties can be modified by surfactant adsorption at the gas-liquid interface. Adsorption of surfactant influences colloidal dispersion by altering the electrostatic interaction, van der Waals interaction, and steric interaction between two NBs. The degree of surface property change depends on the surfactant dissociation, nature of the charge of the head group of the surfactant, adsorption density, and orientations of surfactant molecules at the gas-liquid interface. CTAB is a long-chain molecule with a +ively charged hydrophilic head and a neutral hydrophobic tail.[91] On the other hand, SDS has a −ively charged hydrophilic head and a neutral hydrophobic tail. While generating the NBs in the surfactant solution, the hydrophobic tail, mainly the hydrocarbon chain, aligns towards the gas phase while the hydrophilic head group is towards the bulk liquid phase.[92] Surfactant molecules are almost ionized in the liquid medium. Nevertheless, the adsorption of ionized ions at the NBs surface is limited by the degree of dissociation ($\beta$). The degree of dissociation for SDS and CTAB at the NBs surface as a function of different surfactant concentrations is presented in Figure 9a. The surfactant concentrations range from 0.01 to 2 times the CMC. As observed, the degree of dissociation of the surfactant slightly reduces as the surfactant concentration approaches the CMC. The reduction in the $\beta$ is in the range from 0.98 to 0.96 and from 0.99 to 0.98 for SDS and CTAB, respectively. A slightly high-value of $\beta$ in CTAB indicates the preferential adsorption on NBs surface. The order of adsorption of a cationic, anionic, and nonionic surfactant on the cellulose–water interface was: cationic > anionic $\approx$ nonionic, as reported by another researcher.[93] Above the CMC, a significant reduction in the $\beta$ of both the surfactants can be witnessed. Surfactant molecules begin to form micelles rather than dissociate at CMC and higher concentrations, reducing the $\beta$. When the surfactant concentration increases above CMC, the molecules begin to move towards the bulk rather than being adsorbed at the surface.[94] Figure 9b depicts the theoretical and experimental surface potential ($\psi_0$) as a function of SDS concentration. Theoretical values are determined from Eqs. (7) and (8), whereas experimental values are enunciated using Eq. (13). The comparison suggests a good agreement between theoretical and experimental values.

The adsorption of surfactants on the NBs surface as a function of surfactant concentration is demonstrated in Figure 9c. The number of adsorbed molecules/nm$^2$ increases as the SDS concentration increases to 5.74 mM. The maximum surface coverage is noticed at a surfactant concentration of 5.74 mM. The adsorption curve drops above 5.74 mM surfactant concentration until the CMC is reached. The reduction in the adsorption may be ascribed to the electrostatic hindrance, which is due to the same charge of the surfactant species and the NB surface. The adsorption of the surfactant molecules does not alter above the CMC, which is attributed to the micelle formation in the bulk or monolayer coverage. A similar trend in the adsorption behavior of CTAB, a +ively charged surfactant on −ively charged NB, is shown in the same plot. Adsorption reaches its maximum surface coverage at $c_s = 0.28 \, \text{mM}$ and decreases significantly after this point until the CMC is reached. No changes were witnessed in the adsorption behavior at or above the CMC.

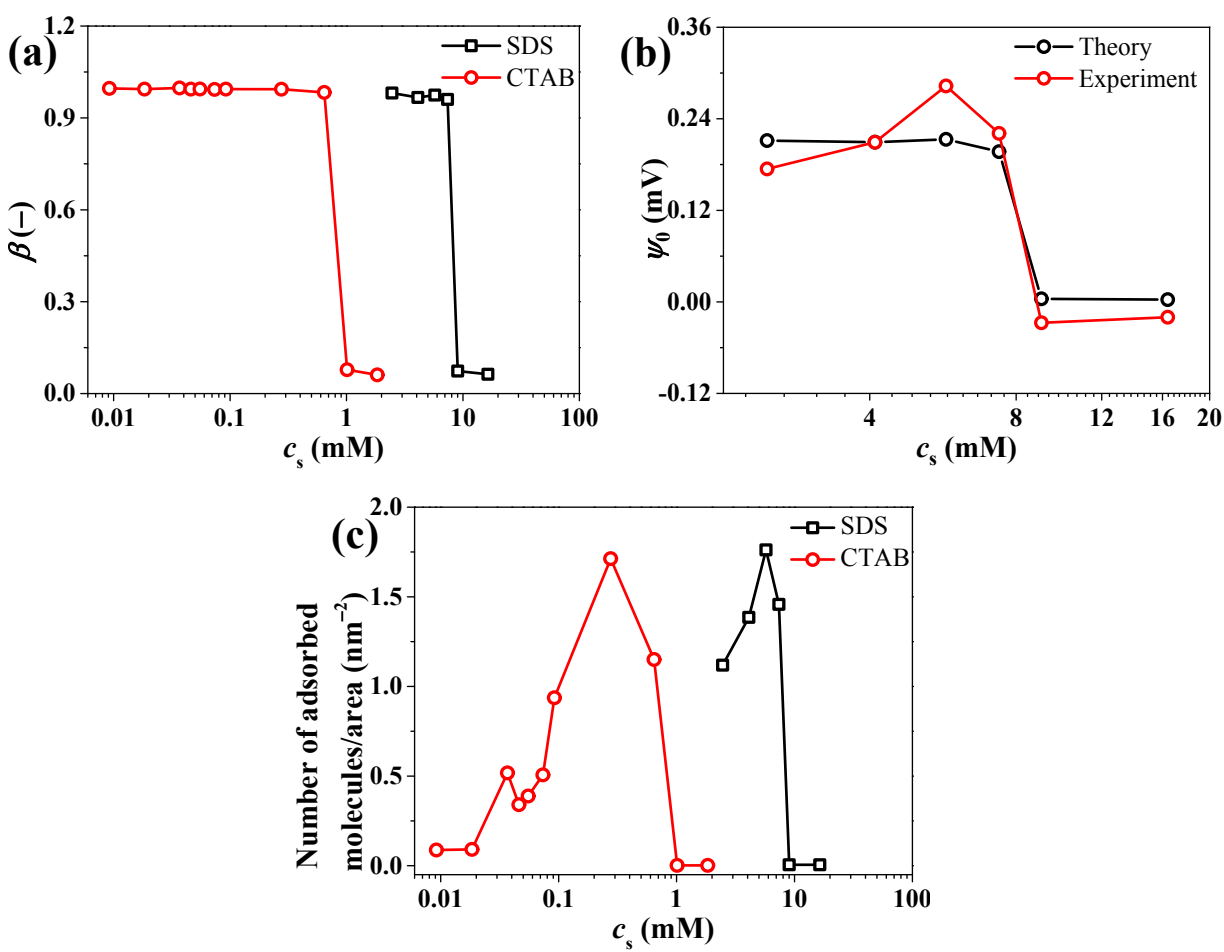

Figure 9. (a) Surfactant dissociation at the NBs surface, (b) comparison of theoretical and experimental surface potential, and (c) number of adsorbed molecules/nm² of NBs surface in ionic surfactants.

## 4.8. Distribution of ions at the gas-liquid interface

In surfactant solutions, the surface charge of bubbles is highly influenced by the kind of surfactant and its ionization characteristics. As per our erudition, the distribution of ions and electric double layer in the presence of surfactants around the NBs has not been elucidated in the literature. A plausible prospect of ion distribution and its alignment around NBs in cationic and anionic surfactants is explicated. Two theories are available to discuss the distribution of the ions on the gas-liquid interface. The first theory states that the counterions of surfactant may react with its head.[95] The second theory neglects the binding processes while including the dipole moment of the adsorbed surfactant molecules on the total potential change.[96] The second theory is more acceptable for ionic surfactants and this theory requires Gibbs adsorption expression to quantify the concentration of surfactant adsorption, as discussed in earlier section. As long as the dissolved surfactants form ionic species, it is −ive for anionic surfactant solutions and +ive for cationic surfactant solutions.

Ions distribution at the gas-liquid interface is demonstrated in Figure 10. In pure water, the $H^+$ ion generated by the hydrolysis of water molecules is mostly free in solution; nevertheless, $OH^-$ often binds at the gas-liquid interface.[97] NBs are −ively charged in pure water, mainly because of excess $OH^-$ ions at the gas-liquid interface and attract counterions with the opposite charge ($H^+$). At the gas-liquid interface, the water molecule dipole orientation and $H^+$ and $OH^-$ hydration energies are used to bind $OH^-$ ions selectively.[98] For instance, $H^+$ has an energy of around − 1127 kJ/mol, whereas $OH^-$ has an energy of approximately − 489 kJ.[71] The surface charge affects the spatial distribution of ions or molecules in the solution. Ions with opposite charges are drawn to the surface, while ions with the same charge are repelled. As a result, a diffused electrical double layer developed, consisting of the charged surface, neutralizing counterions, and co- and counterions distributed far from the surface.[99] The stern layer is the first layer seen in Figure 10, followed by the diffuse layer. The slipping plane is the outer edge of the diffuse layer that separates it from the bulk medium.

In the DI water case, the stern layer mainly consists of $H^+$ ions (counterions) because of the coulombic attraction of $OH^-$ ions, as shown in Figure 10a. The $H^+$ ions will move closer to the interface. In the immediate vicinity of the interface, these ions are closely bonded. The diffuse layer has both $H^+$ ions and $OH^-$.

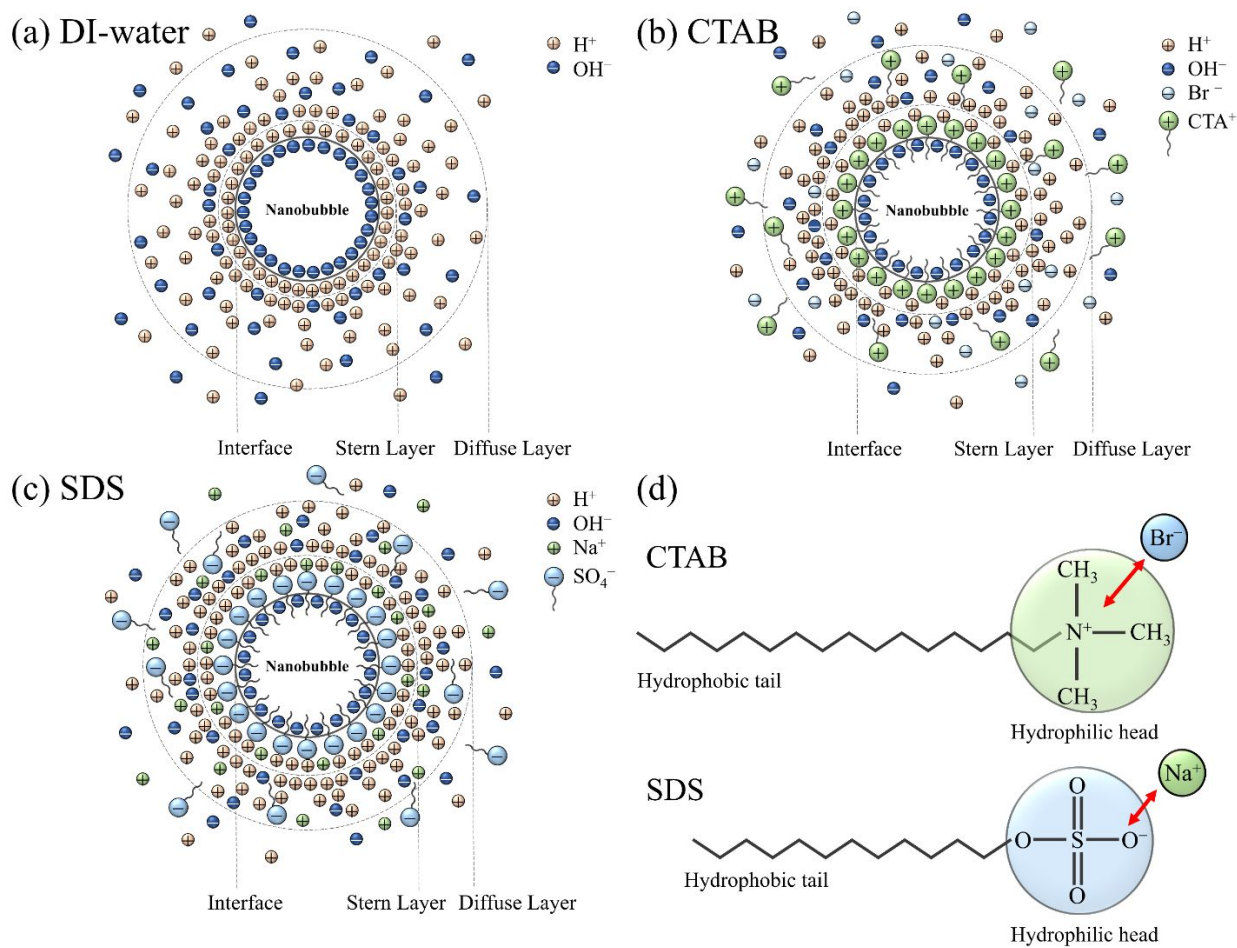

Figure 10. Distributions of ions at the gas-liquid interface in the presence of (a) SDS, (b) CTAB, (c) SDS, and (d) surfactants hydrophilic head and hydrophobic tail.

Ionic surfactant is composed of both hydrophilic and hydrophobic parts, denoted by polar and non-polar groups. The non-polar group is oriented toward the air phase during the adsorption of surfactant molecules, whereas the polar group is oriented towards the water.[100] Due to the differing orientations of polar and non-polar groups of a surfactant, an electric double layer forms at the gas-liquid interface.[38] CTAB possesses a +ively charged hydrophilic head, CTA$^+$, and a neutral hydrophobic tail. There is a surfactant layer at the gas-liquid interface with a neutral hydrophobic tail oriented towards the gas phase and a +ively charged hydrophilic head, CTA$^+$ on the surface, as shown in Figure 10b. The +ively charged hydrophilic head CTA$^+$ dominates the stern layer, while the diffuse layer contains both Br$^-$ and +ively charged hydrophilic heads CTA$^+$. In the bulk phase (after the diffuse layer), the +ively charged hydrophilic head CTA$^+$ would be oriented towards the bulk liquid phase, while the neutral hydrophobic tail will be towards the gas phase.

Slipping plane exists in the diffuse layer and its exact position is a matter of research. Ions that extend beyond the slipping plane will not move as an entity with the NB and the ions are diffused more freely around the NB. Ions inside this boundary will move as one entity with NBs. In aqueous solution, SDS dissociates itself as: $CH_3(CH_2)_{11}SO_4Na = Na^+ + CH_3(CH_2)_{11}SO_4^-$. In the stern layer, −ively charged hydrophilic head $SO_4^-$ groups are adsorbed to the NBs surface, while the neutral hydrophobic tail is aligned towards the gas phase (Figure 10c). The diffuse layer contains $Na^+$ ions and −ively charged hydrophilic $SO_4^-$ groups. It should be underlined that the given ion being categorized in the stern or diffuse layer is still an unresolved query. There is no agreement and conclusive findings are available to support it.

## 5. Conclusions

The present work quantifies the behavior of NBs in DI water, pH medium, and different surfactants such as cationic, anionic, and nonionic. The NBs behavior was analyzed in terms of average bubble size, concentration, bubble size distribution, long-term stability, and surfactant adsorption. NBs were produced using a piezoelectric transducer. Based on the experimental findings, the following conclusion is drawn:

- Zeta potential of NBs was +ive below pH < 4.5, while it was −ive for 4.5 < pH < 11. In DI water, the $\zeta$ was − 25.41 mV. Reducing pH below 7 diminishes the −ive magnitude of $\zeta$, and at pH = 4, the $\zeta$ is reversed from −ive to +ive. The IEP was at pH = 4.5. The reduction in the −ive magnitude of $\zeta$ in an acidic medium was due to the preferential adsorption of $H^+$ ions. The increase in CTAB concentration reduces the magnitude of its −ive $\zeta$. The reduction is due to the favored adsorption of the $CTA^+$ ion on the NBs surface. The charge reversal occurs at a surfactant concentration of $1.84 \times 10^{-2}$ mM. The maximum value of the $\zeta$ was 47.02 mV at a $64.4 \times 10^{-2}$ mM concentration. In the SDS solution, $\zeta$ values were −ive at all concentrations. The −ive $\zeta$ is because of $SO_4^-$ ions. The maximum value of $\zeta$ was − 51.63 mV. Tween 20 does not contain any charge; thus, its presence has little effect on the $\zeta$ of NBs.

- The pH results show that the higher the magnitude of the $\zeta$, the smaller the average bubble size. The smallest bubble size corresponds to its largest magnitude, either −ive or +ive $\zeta$.

The average NBs size in DI water and basic medium is lower than in acidic medium. The presence of surfactants significantly impacts the NBs concentration and size. Average NBs size was observed to decrease with an increase in the ionic surfactant concentration up to the CMC. However, it increases above the CMC. In the nonionic surfactant, a trend of slightly increasing NBs size is observed by increasing the surfactant concentration, and the increase in the size is substantial above the CMC.

- Bubble size distribution (BSD) in DI water is aligned to the smaller bubble size compared to the other pH mediums. BSD at pH = 5 (close to IEP) was perceived to have a larger bubble size than other pH mediums, which is attributed to instability in the surface charge of the bubble that leads to bubble coalescence. BSD curves shift to the small bubble size as pH reduces from 6 to 4 and to 3. In an alkaline medium, the BSD curve is just the right-hand side of pH = 7, attributing small NBs compared to the acidic medium. BSD curves shift to the small bubble size as CTAB concentrations exceed $3.68 \times 10^{-2}$ mM. No significant variation in BSD curves was observed in the presence of SDS. It is due to the no change in the surface charge of NBs from –ive to +ive or vice versa. No particular trend in the BSD curve is witnessed for Tween 20, and several relative frequency peaks can be detected for BSD curves.

- Extended DLVO theory was used to analyze the behavior of NBs based on van der Waals interaction potential, electrostatic repulsive interaction potential, and hydrophobic interaction potential. A low value of the potential energy barrier indicates van der Waals interaction potential dominance compared to electrostatic repulsive interaction potential. For $6 \leq$ pH $\leq 11$, the potential energy barrier is +ive and greater than $43.90k_{\mathrm{B}}T$, whereas for pH < 6, the potential energy barrier almost vanishes. At higher potential energy barriers, NBs are more stable.

- NBs average bubble size and concentrations were examined over days 1 to 76. Over time, a reduction in bubble size is noted for all the pH. No NBs were observed beyond day 18 at a low pH = 3, indicating that it is unsuitable for long-term stability. NBs in surfactant and pH solutions have existed for a long time. The existence of NBs over a long time reflects the low impact of bubble coalescence and Smoluchowski ripening. The persistence of the NBs surface charge and its magnitude over a long time governed the bubble coalescence, hence the stability.

- NBs surface properties can be modified by surfactant adsorption at the gas-liquid interface. The adsorption of ionic surfactants on the NBs surface was analyzed using the Gibbs adsorption expression, and in addition, the degree of dissociation of the surfactant and the number of adsorbed surfactant molecules on the NBs surface were also enunciated. The degree of dissociation reduces slightly as the surfactant concentration increases until it reaches the CMC. Above the CMC, a significant reduction in the degree of dissociation was observed.

- The number of adsorbed molecules increases as the surfactant concentration increases to the CMC. The reduction in the adsorption was noticed above the CMC. The reduction is ascribed to the electrostatic hindrance due to the same charge of the surfactant species and the NB surface. The adsorption of the molecules does not alter above the CMC, which is attributed to the micelle formation in the bulk or monolayer coverage.

- A plausible prospect of ion distribution and its alignment around NBs in an ionic surfactant is explicated.

The findings from this research can give insight into further understanding, process intensification, and modeling and analyzing the peculiar qualities of NBs, such as their size, concentration, and long-term stability in various pH media and ionic and nonionic surfactants.

## ACKNOWLEDGEMENTS

This research was supported by the Basic Science Research Program through the National Research Foundation of Korea (NRF) funded by the Ministry of Education (NRF2021R1A6A1A03039696) and the Basic Science Research Program through the National Research Foundation of Korea (NRF) grant funded by the Ministry of Science, ICT & Future Planning (NRF2020R1A2C3010568). We are also grateful for the support of the Korea Environment Industry & Technology Institute (KEITI) through its Ecological Imitation-based Environmental Pollution Management Technology Development Project, funded by the Korea Ministry of Environment (MOE) (2019002790003).